\documentclass{elsart}

\usepackage{amsmath,amssymb,graphicx,bm}

\newcommand{\openone}{\leavevmode\hbox{ \small1\normalsize\kern-.33em1}} 
\newcommand{\antisymV}{\mbox{\boldmath $\mathcal{V}$}}
\newcommand{\rhobold}{\mbox{\boldmath $\rho$}}
\newcommand{\beqn}{\begin{equation}}
\newcommand{\eeqn}{\end{equation}}
\newcommand{\beq}{\begin{equation}}
\newcommand{\eeq}{\end{equation}}
\newcommand{\bea}{\begin{eqnarray}}
\newcommand{\eea}{\end{eqnarray}}
\newcommand{\ba}{\begin{align}}
\newcommand{\ea}{\end{align}}
\newcommand{\vlowk}{V_{{\rm low}\,k}}
\newcommand{\vsrg}{V_{\rm srg}}

\newcommand{\la}{\Lambda}
\newcommand{\kf}{k_{\text{F}}}

\newcommand{\fmi}{\, \text{fm}^{-1}}
\newcommand{\mev}{\, \text{MeV}}
\newcommand{\gevi}{\, \text{GeV}^{-1}}

\newcommand{\bDel}{\bm{\Delta}}
\newcommand{\bSig}{\bm{\Sigma}}
\newcommand{\Mstar}{M^{\ast}}
\newcommand{\bfx}{{\bf x}}
\newcommand{\xvec}{{\bf x}}
\newcommand{\Gamint}{{\Gamma}_{\rm int}}

\newcommand{\kvec}{{\bf k}}
\newcommand{\kpvec}{{\bf k'}}
\newcommand{\qvec}{{\bf q}}
\newcommand{\pvec}{{\bf p}}
\newcommand{\qb}{\bar{q}}
\newcommand{\pb}{\bar {p}}
\newcommand{\rvec}{{\bf r}}
\newcommand{\rone}{{\bf r}_1}
\newcommand{\rtwo}{{\bf r}_2}
\newcommand{\rthree}{{\bf r}_3}
\newcommand{\rfour}{{\bf r}_4}
\newcommand{\xone}{{\bf x}_1}
\newcommand{\xtwo}{{\bf x}_2}
\newcommand{\xthree}{{\bf x}_3}

\newcommand{\svec}{{\bf s}}

\newcommand{\Rvec}{{\bf R}}

\newcommand{\bfcdot}{\bm{\cdot}}
\newcommand{\imax}{i_{{\rm max}}}
\newcommand{\WHF}{W_{\rm HF}}
\newcommand{\Wint}{W_{\rm int}}
\newcommand{\rhoSL}{\rho_{\rm SL}}
\newcommand{\rhoNM}{\rho_{\rm NM}}

\newcommand{\tr}{{\rm Tr}}

\begin{document}

\begin{frontmatter}

\title{Density Matrix Expansion  \\
       for Low-Momentum Interactions}

\author[OSU,MSU]{S.K.\ Bogner},
\ead{bogner@nscl.msu.edu}
\author[OSU]{R.J.\ Furnstahl\corauthref{cor}},
\corauth[cor]{Corresponding author.}
\ead{furnstahl.1@osu.edu}
\author[OU,OSU]{L. Platter}
\ead{lplatter@mps.ohio-state.edu}
\address[OSU]{Department of Physics,
The Ohio State University, Columbus, OH\ 43210}
\address[MSU]{National Superconducting Cyclotron Laboratory and Department 
of Physics and Astronomy,
Michigan State University, East Lansing, MI 48824}
\address[OU]{Department of Physics and Astronomy,
Ohio University, Athens, OH\ 45701}

\date{\today}

\begin{abstract}
A first step toward a universal nuclear energy density functional
based on low-momentum interactions is taken using
the density matrix expansion (DME) of Negele and Vautherin.
The DME is
adapted
for non-local momentum-space potentials and generalized to include
local three-body interactions.
Different prescriptions for the three-body DME are compared.
Exploratory results are given at the Hartree-Fock level, along with a
roadmap for systematic improvements within an effective action framework
for Kohn-Sham density functional theory.
\end{abstract}

\end{frontmatter}


\section{Introduction}
  \label{sect:introduction}
  
Calculating the properties of atomic nuclei from microscopic
internucleon interactions is one of the most challenging and enduring
problems of nuclear physics.  However, recent
developments in few- and many-body physics together
with advances in computational technology give hope 
that controlled calculations of medium and heavy nuclei
starting from a microscopic nuclear
Hamiltonian 
will be forthcoming (see, for example, 
\cite{recentprogress,CD2006,Furnstahl:2008df}). 
Density functional theory (DFT), which
is a self-consistent framework that goes beyond conventional
mean-field approaches, offers particular promise for medium to 
heavy nuclei.
The central object in DFT is an energy functional of the
nuclear densities that would apply to all the nuclides. 
Phenomenological functionals have had many successes but lack
a microscopic foundation and theoretical control of errors,
such that extrapolations to the limits of nuclear binding are uncontrolled. 

Recent progress in evolving chiral effective field theory (EFT) interactions  
to lower momentum using renormalization group (RG) 
methods
\cite{Vlowk1,Vlowk2,VlowkRG,Vlowk3N,Vlowknm,%
Bogner:2006tw,Bogner:2006pc,Bogner:2007jb,SRG3body}
(see also \cite{Roth:2005pd,rothRPA})
makes feasible a microscopic calculation of a universal nuclear
energy density functional (\hbox{UNEDF})~\cite{UNEDF}.
The evolution weakens or largely eliminates 
sources of non-perturbative
behavior in the two-nucleon sector
such as strong short-range repulsion and the tensor 
force from iterated pion exchange~\cite{Bogner:2006tw}, 
and the consistent three-nucleon interaction
is perturbative at lower cutoffs~\cite{Vlowk3N}.
When applied to nuclear matter, many-body perturbation theory 
for the energy appears
convergent (at least in the particle-particle channel), with
calculations that include most of the second-order contributions
exhibiting saturation in nuclear matter and 
showing relatively weak dependence on
the cutoff \cite{Vlowknm}.
These features are favorable ingredients
for a microscopic Kohn-Sham DFT treatment~\cite{DREIZLER90,ARGAMAN00,fi03}.
Indeed, Hartree-Fock is a reasonable (if not fully quantitative)
starting point, which
suggests that the theoretical developments and phenomenological
successes of DFT for
Coulomb interactions may be applicable to the nuclear case for
low-momentum interactions.

A formal constructive framework for Kohn-Sham
DFT based on effective
actions of composite operators can be carried out using the inversion 
method \cite{FUKUDA,VALIEV,VALIEV97,Polonyi2001,Puglia:2002vk,%
Bhattacharyya:2004qm,Bhattacharyya:2004aw,Furnstahl:2004xn}.
This is an organization of the many-body problem
that is based on calculating the response of a finite 
system to external, static
sources rather than seeking the many-body wave function.
It requires a tractable expansion (such as an EFT momentum expansion
or many-body perturbation theory) that is controllable in the
presence of inhomogeneous
sources, which act as single-particle potentials.
This is problematic for conventional internucleon interactions,
for which the single-particle potential needs to be tuned
to enhance the convergence of the hole-line expansion \cite{DAY67,Baldo99},
but is ideally suited for low-momentum interactions.
Given an expansion, one can construct a free-energy
functional in the presence of the sources
and then Legendre transform order-by-order
to the desired functional of the densities.
However, these are complicated, non-local
functionals and we require functional
derivatives with respect to the densities, whose dependences are
usually only implicit.
While this is a feasible program,
it will require significant
development to extend existing phenomenological
nuclear DFT computer codes. 

We seek a path that will be compatible in the short term
with current nuclear DFT
technology but testable and systematically improvable.
In this regard, 
the phenomenological nuclear energy density functionals of
the Skyrme form have the closest connection to 
low-momentum interactions.  
Modern Skyrme functionals have been applied over a very wide range
of nuclei, with quantitative success in reproducing properties
of nuclear ground states and low-lying 
excitations~\cite{Dobaczewski:2001ed,Stoitsov:2003pd,BENDER2003}.
Nevertheless, a significant reduction of the global and local errors
is a major goal~\cite{SCIDAC}.
One strategy is to improve the functional itself;
the form of the basic Skyrme functional in use is very restricted,
consisting of a sum of local powers of various nuclear
densities [e.g., see Eq.~(\ref{eq:ESHF})].
Fits to measured nuclear data 
have given to date only limited constraints on possible
density and isospin dependences and on the form of the spin-orbit interaction.
Even qualitative insight into these properties from realistic microscopic
calculations should be beneficial
in improving the effectiveness of the energy density functional.

A theoretical connection of the Skyrme functional to free-space NN
interactions was made long ago by Negele and Vautherin using the
density matrix expansion (DME) \cite{NEGELE72,NEGELE75,Hofmann98}, 
but there have
been few subsequent microscopic developments.
The DME originated as an expansion of the Hartree-Fock energy
constructed using the nucleon-nucleon (NN) 
G matrix  \cite{NEGELE72,NEGELE75}, which was
treated in a local (i.e., diagonal in coordinate representation) approximation.
In this paper, we revisit the DME
using non-local low-momentum interactions in momentum representation, 
for which G matrix summations are
not needed because of the softening of the interaction.
When applied to a Hartree-Fock energy functional, the DME yields
an energy functional in
the form of a generalized Skyrme functional that is compatible
with existing codes, by replacing 
Skyrme coefficients with density-dependent functions.
As in the original application, a key feature of the DME is
that it is not a pure short-distance expansion but includes
resummations that treat long-range pion interactions correctly
in a uniform system.
However, we caution that
the Negele-Vautherin DME involves prescriptions for the resummations without
a corresponding power counting to justify them. 

The idea of using soft, non-local potentials in an expansion starting
with Hartree-Fock was explored in the late sixties and
early seventies (see, for example, Refs.~\cite{KERMAN66,KERMAN67,KERMAN73}).
However,
soft potentials were generally abandoned because of their inability to 
saturate nuclear matter at the empirical density and energy per 
particle.%
\footnote{Calculations using hard NN-only interactions also fail to reproduce 
empirical saturation properties.}
They have been revived in the context of low-momentum potentials
(often referred to as ``$\vlowk$'')
derived by transforming modern realistic NN potentials. 
The key to their success is the recognition that three-body forces
(and possibly four-body forces)
cannot be neglected.  With lowered cutoffs, the density dependence of
the three-body contribution drives saturation~\cite{Vlowknm}, 
which accounts for the
apparent past failure in nuclear matter
when only two-body contributions were included.

The present work is 
a proof-of-principle demonstration with a roadmap for future
developments.
We note the following omissions and simplifications.
\begin{itemize}
  \item
We restrict ourselves to isoscalar ($N=Z$) functionals.  This is 
merely for simplicity; generalizations to the full isovector
dependence will be presented in the near future.
We also defer inclusion of spin-orbit and tensor terms, which will 
require extensions of the DME treatment 
of Negele and Vautherin~\cite{Biruk}. 

  \item
We work to leading order in the perturbative many-body
expansion (i.e., Hartree-Fock).  An upgrade path to include
second order and beyond is described in Section~\ref{sec:summary}.

 \item
The form for the three-body force is limited to that of chiral
N$^2$LO EFT.  This is consistent with current approximations
used with low-momentum potentials, but will need to be generalized
to accommodate evolved three-body potentials.

  \item 
Pairing is essential for the quantitative treatment of nuclei,
particularly unstable nuclei.
The DME functionals described here can be adapted to include pairing
as done in conventional Hartree-Fock-Bogliubov phenomenology.
However, a unified treatment is feasible
with low-momentum interactions~\cite{Furnstahl:2006pa,Duguet:2007be}.

  \item
There are unresolved conceptual issues for applying DFT to a
self-bound system  \cite{Engel:2006qu,Giraud:2008zz,Barnea:2007jx}
that we will not address here (but which must
be dealt with eventually).  In addition, projection is not considered.
\end{itemize}

Recently, Kaiser and collaborators have applied the DME in momentum
space to a
perturbative chiral EFT expansion at finite density 
to derive a Skyrme-like energy functional
for nuclei \cite{Kaiser:2002jz,Kaiser:2003uh,Kaiser:2005}.
Their analytic expressions for long-range pion contributions
can be effectively applied in our formalism to avoid slowly
converging partial-wave summations.
However, we defer to future work
a detailed description of this application and also
comparisons with their results.

The plan of the paper is as follows.
In Section~\ref{sec:DFT}, we present the features of density functional theory
needed in our treatment and discuss how applying the DME will lead us
to a generalized Skyrme-like energy functional. 
In Section~\ref{sec:dme-two-body}, we review the Negele/Vautherin derivation of
the DME for non-local (in coordinate space) two-body 
potentials and make a direct
extension to momentum space.  The result is a set of simple formulas for the
basic coefficient functions in terms of integrals over partial-wave
matrix elements of the $\vlowk$ potential.
In Section~\ref{sec:dme-three-body}, we extend the DME to include three-body
forces, restricting ourselves to local potentials of the form used in chiral EFT
at N$^2$LO (which is the form used in current approximations to low-momentum NNN
interactions).
We consider two prescriptions for the three-body part.
We present some tests of the DME and
sample results in Section~\ref{sec:results}, highlighting the effects of
non-locality,
the relative size of NN and NNN contributions, and the impact of
different prescriptions for the NNN DME expansion.
We conclude
with a summary and roadmap for future calculations in
Section~\ref{sec:summary}.


\section{Density Functional Theory}
 \label{sec:DFT}
 
In this section, we give overviews of the standard Skyrme functional
and the ideas behind 
Kohn-Sham DFT for nuclei that we need to set up the energy density
functional calculations using the DME.
 
\subsection{Skyrme Hartree-Fock Energy Density Functional} 
 
In the conventional Skyrme Hartree-Fock (SHF) formalism, 
the energy is a functional of
the density $\rho$, the kinetic density $\tau$, and the spin-orbit
density $\textbf{J}$.
For simplicity, we restrict the discussion to
$N=Z$ nuclei here, so these are isoscalar
densities only.
This functional is a single integral of a local energy density,
which depends in a simple way on these densities, such as~\cite{RINGSCHUCK}
\bea
  E_{\rm SHF}[\rho,\tau,\textbf{J}] 
    &=& \int\!d^3x\,
    \biggl\{ \frac{1}{2M}\tau + \frac{3}{8} t_0 \rho^2
  + \frac{1}{16} t_3 \rho^{2+\alpha}
 + \frac{1}{16}(3 t_1 + 5 t_2) \rho \tau  \nonumber
  \\ & & \hspace*{-.1in}\null
  + \frac{1}{64} (9t_1 - 5t_2) (\bm{\nabla} \rho)^2  
  - \frac{3}{4} W_0 \rho \bm{\nabla}\bfcdot\textbf{J}
  + \frac{1}{32}(t_1-t_2) \textbf{J}^2 \biggr\}  
  \;.
  \label{eq:ESHF}
\eea
Expressions for the Skyrme functional including isovector
and more general densities can be found in Ref.~\cite{PERLINSKA04}.
The densities  $\rho$, $\tau$, and $\bm{J}$ are expressed as sums 
over single-particle orbitals $\phi_\beta(\xvec)$: 
\bea
  \rho(\xvec) &\equiv& \sum_\beta |\phi_\beta(\xvec)|^2 \;,
  \label{eq:rhoeq}
  \\
  \tau(\xvec) &\equiv& \sum_\beta |\bm{\nabla}\phi_\beta(\xvec)|^2
  \;,
  \label{eq:taueq} 
  \\
  \textbf{J}(\xvec) &\equiv& \sum_\beta \phi^\dagger_\beta(\xvec)
    (-i\bm{\nabla\times\sigma})\phi_\beta(\xvec)
  \;,
  \label{eq:skyrmeeqs}
\eea
where the sums are over occupied states and the spin-isospin
indices are implicit.
(More generally, when pairing is included with a zero-range interaction, 
the sums are over all
orbitals up to a cutoff, weighted by pairing occupation numbers.
This complicates finding the self-consistent solution significantly
but is not important for our discussion.) 
The parameters
$t_0$--$t_3$, $W_0$, and $\alpha$ 
determine the functional and
are obtained from numerical fits to
experimental data.

Varying the energy with
respect to the wavefunctions with Lagrange multipliers $\varepsilon_\beta$
to ensure normalization%
\footnote{Unconstrained variation of the orbitals 
is the usual textbook formulation of Skyrme Hartree-Fock~\cite{RINGSCHUCK}.
But this \emph{does not} hold  beyond Hartree level
for a general microscopic DFT treatment with
finite-range potentials, for which there is 
an additional constraint to the orbital variation \cite{fi03}.} 
leads to a Schr\"odinger-type equation 
with a position-dependent mass term \cite{VB72,RINGSCHUCK}:
\beqn
  \Bigl( - \bm{\nabla} \frac{1}{2M^*(\xvec)} \bm{\nabla} + U(\xvec) + 
   \frac{3}{4} W_0\bm{\nabla} \rho\cdot \frac{1}{i}
   \bm{\nabla} \times \bm{\sigma}
   \Bigr)\,
  \phi_\beta(\xvec) =
  \varepsilon_\beta\,\phi_\beta(\xvec)\, ,\label{eq:SkyrmeEq}
\eeqn
where~\cite{RINGSCHUCK} 
\beqn
   U(\xvec) = \frac{3}{4}t_0\rho + \frac{3}{16}t_3\rho^2
     + \frac{1}{16}(3t_1 + 5 t_2)\tau
     + \frac{1}{32}(5t_2 - 9 t_1) \bm{\nabla}^2\rho
     - \frac{3}{4}W_0 \bm{\nabla\cdot}\textbf{J}
     \;,
     \label{eq:Ueq}
\eeqn
the effective mass $M^*(\xvec)$ is 
\beqn
  \frac{1}{2M^{*}(\xvec)}
   =\frac{1}{2M} + \left[\frac{3}{16}\,t_1
       +\frac{5}{16}\,t_2\right]\,\rho(\xvec)\,
       ,\label{eq:MassEq}
\eeqn
and the $W_0$ term is a spin-orbit potential (see Ref.~\cite{JACEK95} for
details).
The potentials in 
Eqs.~(\ref{eq:Ueq})--(\ref{eq:MassEq}) and the orbitals from
Eq.~(\ref{eq:SkyrmeEq}) are evaluated alternately until 
self-consistency (see Fig.~\ref{fig:dft_diagrams}).

As we will see below,
the DME energy functional for $N=Z$ will take the same local form
as $E_{\rm SHF}$,
\beqn
   E_{\rm DME}[\rho,\tau,\textbf{J}] 
     = \int\!d^3R\, \mathcal{E}_{\rm DME}(\rho(\Rvec),\tau(\Rvec),
                \textbf{J}(\Rvec))
     \;,
     \label{eq:EDME}
\eeqn
where the energy density function $\mathcal{E}_{\rm DME}$ is evaluated
with the local densities at $\Rvec$.
We follow the Negele/Vautherin notation for $\mathcal{E}_{\rm DME}$ 
and write~\cite{NEGELE72}
\beqn
 \mathcal{E}_{\rm DME} = \frac{\tau}{2M}
       + A[\rho] + 
        B[\rho]\tau + 
        C[\rho]|\bm{\nabla}\rho|^2 + \cdots
   \;,
   \label{eq:eDME}     
\eeqn 
where $A$, $B$, $C$ are functions of the isoscalar density $\rho$
instead of the constant Skyrme parameters, and we have suppressed
terms that go beyond the present limited discussion. 
(When $N\neq Z$, these are functions of the isovector densities
as well.) 
Equation~(\ref{eq:eDME}) implies that
the DME form will be a direct generalization of the Skyrme functionals.

\subsection{DFT from Effective Actions}
  \label{subsec:effact}

Microscopic DFT follows from calculating the response of a many-body system
to external sources, as in Green's function methods, only with local, static
sources that couple to densities rather than fundamental fields.  
(Time-dependent sources can be used for certain excited states.)
It is profitable to think in terms of a thermodynamic formulation of
DFT, which uses the effective action formalism \cite{NEGELE88}
applied to composite operators 
to construct energy
density functionals \cite{FUKUDA,VALIEV,Polonyi2001}. 
The basic plan is to consider the zero temperature limit
of the partition function $\mathcal{Z}$ for
the (finite) system of interest in
the presence of external sources coupled to various quantities of
interest (such as the fermion density).
We derive energy functionals of these quantities by Legendre
transformations with respect to the sources \cite{Kutzelnigg}. 
These sources probe, in a
variational sense, configurations near the ground state. 

An analogous system would be a lattice of interacting spins, to which we 
apply an
external source in the form of a magnetic field $H$ \cite{NEGELE88}.  
The Helmholtz free energy $F[H]$
is calculated as the energy in the presence of the magnetic field
and we determine the magnetization 
by a derivative with respect to the field, $M(H) = -\partial F[H]/\partial H$.
It is often useful to reverse the problem, and ask what external field produces
a specified magnetization.
This leads us to the Gibbs free energy $G[M]$, which we obtain by
inverting $M(H)$ to find $H(M)$ and performing a Legendre transform:
\beqn
  G[M] = F[H] + H(M) M
  \;.
\eeqn
Because $H = \partial G[M]/\partial M$ and $H$ vanishes in the ground state,
$G$ is extremized in the ground state (and concavity
tells us that it is a minimum). 
If $H$ is an inhomogeneous source, the formalism is generalized 
by replacing partial derivatives by functional derivatives and performing
a functional Legendre transform.     

To derive density functional theory,   
we follow the same procedure, but with sources that adjust density
distributions
rather than spins.
(We can either introduce a chemical potential or only consider variations
that preserve net particle number.  We implicitly assume the latter here.)
Consider first the simplest case of a single 
external source $J(\bfx)$ coupled to the density operator 
$\widehat \rho(x) \equiv \psi^\dagger(x)\psi(x)$ in the partition
function
\beqn
    \mathcal{Z}[J] = 
    e^{-W[J]} \sim {\rm Tr\,} 
      e^{-\beta (\widehat H + J\,\widehat \rho) }
    \sim \int\!\mathcal{D}[\psi^\dagger]\mathcal{D}[\psi]
    \,e^{-\int\! [\mathcal{L} + J\,\psi^\dagger\psi]} 
    \;,
\eeqn
for which we can construct a path integral representation
with Lagrangian $\mathcal{L}$ \cite{NEGELE88}.
(Note: because our treatment is schematic, for convenience
we neglect normalization factors and take the inverse temperature
$\beta$ and the volume $\Omega$ equal to unity in the sequel.)
The static density $\rho(\bfx)$ in the presence of $J(\bfx)$ is
\beqn
  \rho(\bfx) \equiv \langle \widehat \rho(\bfx) \rangle_{J}
   = \frac{\delta W[J]}{\delta J(\bfx)}
   \;,
\eeqn  
which we invert to find $J[\rho]$ and then Legendre transform from $J$ to
$\rho$:
\beqn
   \Gamma[\rho] = - W[J] + \int\!d^3x\, J(\bfx) \rho(\bfx) \;,
   \label{eq:gammarho}
\eeqn
with
\beqn
   J(\bfx) = \frac{\delta \Gamma[\rho]}{\delta \rho(\bfx)}
   \longrightarrow 
   \left.
   \frac{\delta \Gamma[\rho]}{\delta \rho(\bfx)}\right|_{\rho_{\rm gs}(\bfx)
   } =0
   \;.
   \label{eq:Jofx}
\eeqn 
For static $\rho(\bfx)$, $\Gamma[\rho]$ is proportional to 
the conventional Hohenberg-Kohn energy functional, which
by Eq.~(\ref{eq:Jofx}) is extremized at the ground state density
$\rho_{\rm gs}(\bfx)$ (and thermodynamic arguments establish that it is
a minimum \cite{VALIEV97}).\footnote{A Minkowski-space formulation of
the effective action with time-dependent sources
leads naturally to an RPA-like generalization
of DFT that can be used to calculate properties of collective excitations.}   

We still need a way to carry out the inversion from $\rho[J]$ to $J[\rho]$; 
a general approach is
the inversion method of Fukuda et al. \cite{FUKUDA,VALIEV}.
The idea is to expand the relevant quantities in a hierarchy,
labeled by a counting parameter $\lambda$,
\bea
   W[J,\lambda] &\!=\!& W_0[J] + \lambda W_1[J] + \lambda^2 W_2[J] + \cdots 
    \;, \\
   J[\rho,\lambda] &\!=\!& J_0[\rho] + \lambda J_1[\rho] + \lambda^2 J_2[\rho] 
      + \cdots \;, \\
   \Gamma[\rho,\lambda] &\!=\!& \Gamma_0[\rho] 
            + \lambda \Gamma_1[\rho] + \lambda^2 \Gamma_2[\rho] + \cdots
            \;, 
\eea
treating $\rho$ as order unity (which is the same as requiring
that there are no corrections to the zero-order density),
and match order by order in $\lambda$ to determine the
$J_i$'s and $\Gamma_i$'s.  
Zeroth order is a noninteracting system with potential $J_0(x)$:
\beqn
  \Gamma_0[\rho] = -W_0[J_0] + \int\!d^3x\, J_0(\bfx)\rho(\bfx) 
\eeqn
and
\beqn  
  \rho(\bfx) = \frac{\delta W_0[J_0]}{\delta J_0(\bfx)} 
  \;.   
  \label{eq:zeroth}
\eeqn    
Because $\rho$ appears only at zeroth order, 
it is always specified
from the non-interacting system according 
to Eq.~(\ref{eq:zeroth}); there are no corrections at
higher order.
This is the Kohn-Sham system with the same density as the
fully interacting system.

What we have done is to use the freedom to
split $J$ into $J_0$ and $J - J_0$, which is essentially
the same as introducing a single-particle potential $U$
and splitting the Hamiltonian according to 
$H = (H_0 + U) + (V - U)$.  Typically $U$ is chosen to accelerate
(or even allow) convergence of a many-body expansion
(e.g., the Bethe-Brueckner-Goldstone theory \cite{DAY67,RAJARAMAN67,Baldo99}).
For DFT,
we choose it to ensure that the \emph{density} is unchanged, order by order.
Thus, we need the flexibility in the many-body expansion
to choose $U$ without seriously degrading the convergence; 
such freedom is
characteristic of low-momentum interactions.
(Note: If there is a non-zero external potential, it is simply
included with $J_0$.)

We diagonalize $W_0[J_0]$ by introducing Kohn-Sham orbitals $\phi_i$
and eigenvalues $\varepsilon_i$,
\beqn
  [-\bm{\nabla}^2/2m - J_0(\bfx)]\phi_i
  = {\varepsilon_i}\phi_i
  \label{eq:ksequation}
\eeqn
so that
\beqn   
  \rho(\bfx) = \sum_{i=1}^A |\phi_i(\bfx)|^2
   \;.
   \label{eq:ksdensity}
\eeqn
Then $W_0$ is equal to
the sum of $\varepsilon_i$'s.
The orbitals and eigenvalues are used to construct the Kohn-Sham
Green's functions, which are used as the propagator lines in
calculations the $W_i[J_0]$ diagrams.
Finally, we find
$J_0$ for the ground state by truncating the
chain at $\Gamma_{\imax}$,
\beqn
  {J_0} \rightarrow W_1 \rightarrow \Gamma_1 \rightarrow J_1
   \rightarrow W_2 \rightarrow \Gamma_2 \rightarrow \cdots
   \rightarrow W_{\imax} \rightarrow \Gamma_{\imax}
\eeqn
and completing the self-consistency loop:
\beqn
     {J_0(\bfx) = 
     -\sum_{i>0}^{\imax} J_i(\bfx) =
     \sum_{i>0}^{\imax} \frac{\delta\Gamma_{i}[\rho]}{\delta\rho(\bfx)}} 
     \equiv \frac{\delta\Gamma_{\rm int}[\rho]}{\delta\rho(\bfx)}
   \;.   
   \label{eq:loop}
\eeqn
Calculating the successive $\Gamma_i$'s, whose sum is directly
proportional to the desired energy functional, is described in 
Refs.~\cite{VALIEV,VALIEV97,RASAMNY98,Puglia:2002vk}.

When transforming from $W_i$ to $\Gamma_i$,
there are additional diagrams that take into account the adjustment of the
source to maintain the same density and also so-called anomalous diagrams
(these are two-particle reducible).  
A general discussion and Feynman rules for these diagrams are given in
Refs.~\cite{VALIEV,VALIEV97,Puglia:2002vk}.
These two types of contribution cancel up through N$^3$LO in an EFT
expansion
with short-range forces using dimensional
regularization~\cite{Puglia:2002vk}, just as they do
in the inversion method used long ago by
Kohn, Luttinger, and Ward \cite{KOHN60,LUTTINGER60} to show the
relationship of zero-temperature diagrammatic calculations to ones
using the finite-temperature Matsubara formalism in the zero-temperature
limit.
In the present application of the DME approximation to the effective action 
DFT formalism, 
they also cancel and so are omitted entirely.

Note that even though solving for Kohn-Sham orbitals makes the approach
look like a mean-field Hartree calculation, 
the approximation to the energy and density is 
\emph{only} in the truncation of Eq.~(\ref{eq:loop}).
It is a mean-field formalism in the 
sense of a conventional
loop expansion, which is nonperturbative only in the
background field while including further correlations perturbatively
order-by-order in loops.  
The special feature of DFT is
that the saddlepoint evaluation applies the condition that there are no
corrections to the density.
We emphasize that this is not ordinarily an appropriate expansion
for internucleon interactions; it is the special features of
low-momentum interactions that make them suitable.

To generalize the energy functional to accommodate additional
densities such as $\tau$ and ${\bf J}$, 
we simply introduce an additional source coupled to each density. 
Thus,
to generate a DFT functional of the kinetic-energy density
as well as the density, add
$\eta({\bf x})\,\bm{\nabla}\psi^\dagger\bm{\cdot}\bm{\nabla}\psi$ 
to the Lagrangian and
Legendre transform to an effective action of $\rho$ and 
$\tau$~\cite{Bhattacharyya:2004aw}:
\beqn
  \Gamma[\rho,\tau] = W[J,\eta]
   - \int\! d^3x\, J(\bfx)\rho(\bfx) - \int\! d^3x\, \eta(\bfx)\tau(\bfx) 
  \;.
\eeqn
The inversion method results in two Kohn-Sham potentials,
\beqn
     J_0({\bf x}) = 
     \left. \frac{\delta \Gamma_{\rm int}[\rho,\tau]}{\delta
     \rho({\bf x})}\right|_{\tau}
      \quad \mbox{and} \quad 
     \left. \eta_0({\bf x}) = \frac{\delta \Gamma_{\rm int}[\rho,\tau]}{\delta
     \tau({\bf x})}\right|_{\rho} \;,     
\eeqn
where $\Gamma_{\rm int} \equiv \Gamma - \Gamma_0$.
The Kohn-Sham equation is now~\cite{Bhattacharyya:2004aw} 
\beqn
   \bigl[ 
   -\bm{\nabla}{\frac{1}{\Mstar({\bf{x}})}}\bm{\nabla}
     - J_0(\bfx)
   \bigr]\, \phi_i = \epsilon_i \phi_i
   \;,
\eeqn         
with an effective mass
$1/2\Mstar(\bfx) \equiv 1/2M - \eta_0(\bfx)$,
just like in Skyrme HF. 
Generalizing to the spin-orbit or other densities (including
pairing \cite{Furnstahl:2006pa}) proceeds analogously.
We note that the variational principle implies that adding sources will
always improve the effectiveness of the energy functional.

\begin{figure}[t]
 \begin{center}
  \includegraphics*[width=3.8in]{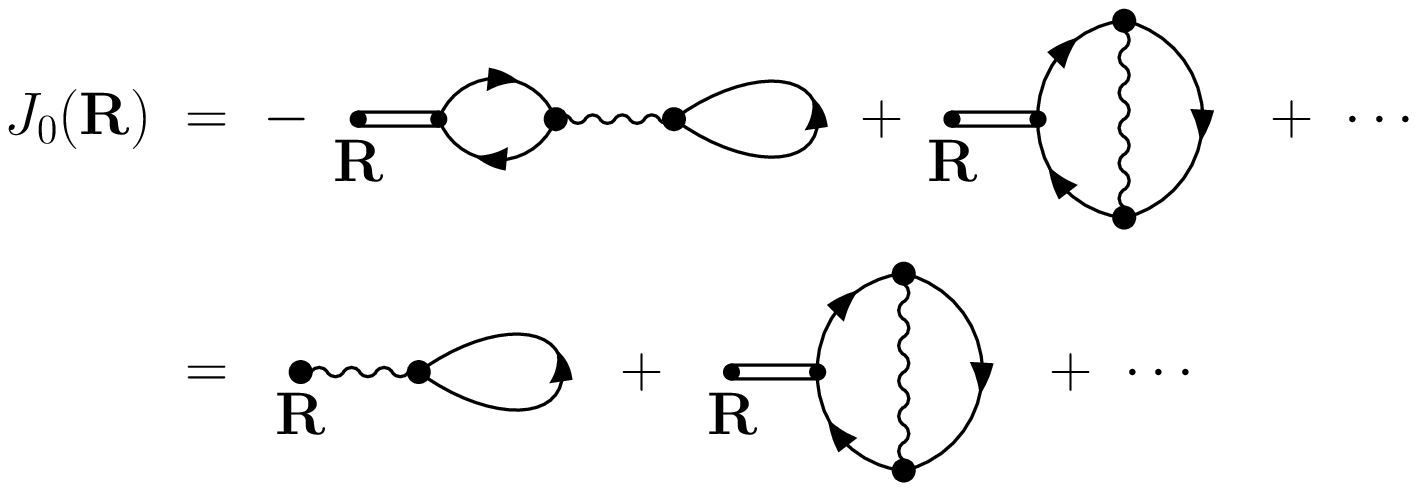}
 \end{center}
 \caption{Schematic representation of Eq.~(\ref{eq:J0chain}) for
 a local potential, where
   the double-line symbol denotes the $(\delta\rho/\delta J_0)^{-1}$ term. }
 \label{fig:J0}
\end{figure}

The Feynman diagrams for $W_i$ will in general include multiple
vertex points over which to integrate.  Further, the dependence on the
densities
will not be explicit except when we have Hartree terms with a 
local potential (that is, a potential diagonal in coordinate
representation).
One way to proceed is to calculate 
the Kohn-Sham potentials
using a functional chain rule, e.g.,
\beqn
 J_0(\Rvec)
  = \frac{\delta\Gamint[\rho]}{\delta\rho(\Rvec)}
  = \int\!d{\bf y}\, \left(
     \frac{\delta\rho(\Rvec)}{\delta J_0({\bf y})}
          \right)^{-1}
   \frac{\delta\Gamma_{\rm int}[\rho]}{\delta J_0({\bf y})}
   \;,
   \label{eq:J0chain}
\eeqn
and steepest descent~\cite{VALIEV97}.
This is illustrated schematically for a local interaction
in Fig.~\ref{fig:J0}.
We see that the Kohn-Sham potential is always just a function
of $\Rvec$ but that the functional is very non-local.
If zero-range interactions are used, these diagrams collapse
into an expression for $J_0(\Rvec)$ that has no internal
vertices, but this is
no longer true for diagrams with more than one interaction.
Orbital-based methods take the chain rule 
in Eq.~(\ref{eq:J0chain}) one step further, adding
a functional derivative of the sources 
with respect to the $\phi_i$'s (and
$\varepsilon_i$'s); see Refs.~\cite{fi03,Ba05,Go05,Ba05b}
for background on these calculations applied to electronic systems.
Eventually, we plan to carry out such calculations to construct the 
full energy density functional.

An alternative in the short term
is to approximate $\Wint$ so that the dependence
on the densities (rather than the sources or the orbitals) 
is explicit.  
This has two effects:  the construction of the $\Gamma_i$ from
the $W_i$ does not have additional terms and
the necessary functional derivatives
are immediate. 
An example of such an approach is the local density approximation (LDA).
Here we go beyond the LDA with the density matrix expansion (DME).
By expanding the $W_i$ about a ``center-of-mass''
$\Rvec$, we generate a local energy density
that is a function of densities ($\rho$, $\tau$, \ldots)
at $\Rvec$.
We choose sources to match these densities
and carry out the Legendre transformation
implicitly; the end result at leading order is calculating $W_1$ 
using density matrices built from Kohn-Sham orbitals.
We are able to vary with respect to the orbitals because the constraint of 
a multiplicative Kohn-Sham potential is built in.
Then the resulting 
Kohn-Sham DFT has precisely the form of the Skyrme Hartree-Fock energy
functional and single-particle equations.

\subsection{Low-Momentum Potentials}

The original DME application was based on a Hartree-Fock 
energy functional calculated with a G matrix,
following the Brueckner-Bethe-Goldstone (BBG) 
method~\cite{DAY67,RAJARAMAN67,Baldo99}.
The latter involves infinite
resummations of diagrams for nuclear many-body theory, as needed
to deal with strongly repulsive potentials.
In BBG there are two general resummations:  the ladder diagrams into
a G matrix and the hole-line expansion using the G matrix.
Furthermore, to accelerate convergence of the hole-line
expansion one needs to carefully
choose a single-particle potential.
This is problematic for the success of a Kohn-Sham DFT construction,
for which the background field (which acts as a single-particle
potential) has a separate constraint, namely to maintain the fermion
density distribution.

Renormalization group (RG) methods can be used to evolve
realistic nucleon-nucleon potentials (e.g., chiral EFT potentials
at N$^3$LO), which typically have strong
coupling between high and low momentum (i.e., off-diagonal matrix
elements of the potential in momentum representation are substantial),
to derive low-momentum potentials in which high and low momentum parts
are decoupled.
This can be accomplished by lowering a momentum cutoff 
$\Lambda$~\cite{Vlowk1,Vlowk2,VlowkRG,Vlowk3N} or
performing a series of unitary transformations that drive the
hamiltonian toward the
diagonal~\cite{Bogner:2006pc,Bogner:2007jb,SRG3body}.
The UCOM transformations of Ref.~\cite{Roth:2005pd} is an alternative to
explicit RG methods.
In all cases, we have a potential for which only low momenta
contribute to low-energy nuclear observables, such as the binding
energy of nuclei.
For convenience, we'll refer to any of
these as $\vlowk$.

We stress that evolving $\vlowk$  does not lose relevant
information for low-energy physics, which includes nuclear ground states
and low-lying excitations, as long as the leading many-body
interactions are kept~\cite{Bogner:2007jb}.
The long-range physics, which is from pion exchange (and Coulomb), is
preserved and remains local, while relevant short-range physics is encoded in the
low-momentum potential through the RG evolution.
Most important, for any $\vlowk$ potential the obstacles from
strongly repulsive potentials are removed.
Hartree-Fock (including three-body interactions)
saturates nuclear matter and G~matrix resummations are
not required (but may still be advantageous).  
Thus, we have a hierarchy suitable for DFT based on many-body
perturbation theory.  [Note: While the need for particle-hole resummations 
remains to be investigated for $\vlowk$ potentials, results from the
analogous UCOM potentials indicate perturbative particle-hole
contributions for the energy~\cite{rothRPA}.]

While the evolution of $\vlowk$ potentials does not disturb the locality
of initial long-range potentials, the short-range part becomes
increasingly non-local. That is, in coordinate representation
$\langle {\bf r} | V | {\bf r'} \rangle$ has an increasing range
in $|{\bf r} - {\bf r'}|$.
Thus we must test that the DME is a good expansion for such
non-localities.

The interactions must include
three-body (and higher-body) potentials, which should be consistently
evolved with the two-body potential.  
These are not yet available (although SRG methods show promise
of providing them in the near 
future~\cite{Bogner:2006pc,Bogner:2007jb,SRG3body}),
and are instead 
approximated by adjusted 
chiral N$^2$LO three-body potentials \cite{Vlowk3N}.  
The validity of this approximation
relies on the RG methods modifying only the short-distance part
of the potential and is supported by the observation that the EFT hierarchy
of many-body forces appears to be preserved by the RG running \cite{Vlowk3N}.
The N$^2$LO three-body 
potentials are local and we restrict our present investigation
for now to this option.
Given this microscopic NN and NNN input, we apply the density matrix
expansion to derive an energy density functional of the Skyrme form.


\section{DME for Two-Body Potentials in Momentum Space}
  \label{sec:dme-two-body}
In this section we derive the density matrix
expansion for a microscopic DFT starting
from low-momentum (and non-local) two-body potentials. 
From Section~\ref{subsec:effact}, 
the relevant object we need to
expand is $\Wint$, which is expressed in terms of 
the Kohn-Sham orbitals and eigenvalues that comprise the Kohn-Sham
single-particle propagators. For Hartree-Fock contributions of the 
form in Fig.~\ref{fig:dme_nonlocal}(a),
however, only the orbitals enter because the Kohn-Sham Green's
function reduces to the density matrix. Similarly, higher-order 
contributions such as the ladder diagrams in the particle-particle (pp)
channel can also be put approximately into this form by averaging
over the state dependence arising from 
the intermediate-state energy denominators. 
Therefore, while the results in this section are derived for the 
Hartree-Fock contributions to the functional, they can easily be
generalized to include higher-order ladder contributions; this
will be explored in a future publication.

In essence, the DME
maps the orbital-dependent expressions for 
contributions to $\Wint$ of the type in 
Fig.~\ref{fig:dme_nonlocal}(a) into 
a quasi-local form, with explicit dependence on the local 
densities $\rho(\Rvec)$, $\tau(\Rvec)$, $\nabla^2\rho(\Rvec)$, 
and so on.
This greatly simplifies the determination of the
Kohn-Sham potential because the functional
derivatives of $\Gamma_{\rm int}$ can be evaluated directly.

\subsection{Expression for $\WHF$}

Before presenting the details of the DME derivation and
its application to non-local low-momentum interactions, it
is useful to first derive in some detail the starting expression
for $\WHF$,
the Hartree-Fock contribution to $\Wint$. This will serve to introduce our basic notation
and to highlight the differences between most existing DME 
studies, which are formulated with local interactions and in coordinate 
space throughout, and the current approach, which is formulated
in momentum space and geared towards non-local potentials.

For a local potential, the distinction between the direct (Hartree) 
and exchange (Fock)
contributions is significant, 
and is reflected in the conventional decomposition
of the DFT energy functional for Coulomb systems, 
which separates out the Hartree piece.
For a non-local potential, the distinction is blurred because
the Hartree contribution now involves the density matrix
(as opposed to the density)
and it is not useful to make this separation when the range
of the interaction is comparable to the non-locality.%
\footnote{However, it is useful to separate out the long-distance
part of the potential, which is local, and treat its direct 
(Hartree) contribution
exactly.}
Consequently, throughout this section we work instead 
with an antisymmetrized interaction. 

For a general (i.e., non-local)
free-space two-body potential $\widehat{V}$, $\WHF$ is defined in terms
of Kohn-Sham states [Eq.~(\ref{eq:ksequation})] labeled by $i$ and $j$, 
\bea
 \WHF &=&\frac{1}{2}
\sum_{i j}^{A} \langle i j | \widehat V(1-P_{12})|i j\rangle 
= \frac{1}{2}\sum_{i j}^{A} \langle i j | \widehat{ \mathcal{V}}|i
j\rangle  \;.
\eea
The summation is over the occupied states
 and the
antisymmetrized interaction $\widehat{\mathcal{V}}=\widehat V (1-P_{12})$  has
been introduced, with the exchange operator $P_{12}$ equal to
the product of operators for spin, isospin, and space exchange,
$P_{12} = P_{\sigma}P_{\tau}P_r$.  Note that the dependence
of $\WHF$ on the Kohn-Sham potential has been suppressed. 
By making repeated use of the completeness relation 
\beqn
 \openone =  \sum_{\sigma\tau}\int\!d{\bf r}|{\bf r}\sigma\tau\rangle\langle 
{\bf r}\sigma\tau|
  \;,
\eeqn
$\WHF$ can be written in
terms of the coordinate space Kohn-Sham orbitals as
\bea
\label{eq:HF_orbitals}
\WHF &=&\frac{1}{2}\sum_{i j} \sum_{\{\sigma \tau\}}
\!\int\! d\rone\! \int\! d\rtwo\! \int\! d\rthree\! \int\! d\rfour\,
\langle \rone\sigma_1\tau_1\rtwo\sigma_2\tau_2|\widehat{\mathcal{V}}
|\rthree\sigma_3\tau_3\rfour\sigma_4\tau_4\rangle
\nonumber
\\
&&\quad\null\times
\phi^*_{i}(\rone\sigma_1\tau_1)\phi_{i}(\rthree\sigma_3\tau_3)
\phi^*_{j}(\rtwo\sigma_2\tau_2)\phi_{j}(\rfour\sigma_4\tau_4) \;.
\eea

From the definition of the Kohn-Sham density matrix,
\beqn
  \label{eq:densitymat}
  \rho(\rthree\sigma_3\tau_3,\rone\sigma_1\tau_1)=
  \sum_{i}^{A}\phi^*_{i}(\rone\sigma_1\tau_1)\phi_{i}(\rthree\sigma_3\tau_3)
  \;,
\eeqn
so Eq.~(\ref{eq:HF_orbitals}) can be written as 
\bea
  \label{eq:HF_DM1}
  \WHF &=&\frac{1}{2} \sum_{\{\sigma \tau\}}
  \!\int\! d\rone\cdots \int\! d\rfour\,
  \langle \rone\sigma_1\tau_1\rtwo\sigma_2\tau_2|\widehat{\mathcal{V}}
  |\rthree\sigma_3\tau_3\rfour\sigma_4\tau_4\rangle
  \nonumber
  \\
  &&\quad\times
  \rho(\rthree\sigma_3\tau_3,\rone\sigma_1\tau_1)
  \rho(\rfour\sigma_4\tau_4,\rtwo\sigma_2\tau_4)\nonumber
  \\
  &=&\frac{1}{2}\tr_{1}\tr_{2}
  \!\int\! d\rone\cdots \int\! d\rfour\, \langle\rone\rtwo| \antisymV^{1\otimes 2} |\rthree\rfour\rangle
  \rhobold^{(1)}(\rthree,\rone)\rhobold^{(2)}(\rfour,\rtwo)
   \;,
\eea
where a matrix notation is used
in the second equation and the traces denote 
summations over the spin and isospin indices 
for ``particle 1" and ``particle 2''. Hereafter we drop the 
superscripts on  $\antisymV$ and $\rhobold$
that indicate
which space they act in as it will be clear from the context.  

Expanding the $\rhobold$ matrices on Pauli spin and isospin matrices we have
\beqn
\rhobold(\rvec_1,\rvec_2) = \frac{1}{4}[\rho_0(\rvec_1,\rvec_2) + \rho_1(\rvec_1,\rvec_2)\tau_z
                       + \vec{S}_0(\rvec_1,\rvec_2)\cdot\vec{\sigma}
           +\vec{S}_1(\rvec_1,\rvec_2)\cdot\vec{\sigma}\tau_z]
	   \;,
\eeqn
where we have assumed the absence of charge-mixing in the single-particle states.
The usual scalar-isoscalar, scalar-isovector, vector-isoscalar, and vector-isovector 
components are obtained by taking the relevant traces,
\bea
 \rho_0(\rvec_1,\rvec_2) &\equiv&
 \tr_{\sigma\tau}[\rhobold(\rvec_1,\rvec_2)] 
 =\sum_{i}^{A}\phi^{\dagger}_i(\rvec_2)\phi_i(\rvec_1) \;,
 \\
 \rho_1(\rvec_1,\rvec_2) &\equiv&
 \tr_{\sigma\tau}[\rhobold(\rvec_1,\rvec_2)\tau_z] =
 \sum_{i}^{A}\phi^{\dagger}_i(\rvec_2)\tau_z\phi_i(\rvec_1) \;,
 \\
 \vec{S}_0(\rvec_1,\rvec_2) &\equiv&
 \tr_{\sigma\tau}[\rhobold(\rvec_1,\rvec_2)\vec{\sigma}] =
 \sum_{i}^{A}\phi^{\dagger}_i(\rvec_2)\vec{\sigma}\phi_i(\rvec_1) \;,
 \\
  \vec{S}_1(\rvec_1,\rvec_2) &\equiv&
 \tr_{\sigma\tau}[\rhobold(\rvec_1,\rvec_2)\vec{\sigma}\tau_z] =
 \sum_{i}^{A}\phi^{\dagger}_i(\rvec_2)\vec{\sigma}\tau_z\phi_i(\rvec_1)\;,
   \label{eq:densitymatrix}
\eea
where $\phi_i(\rvec)$ denotes a spinor with components $\phi_i(\rvec\sigma\tau)$.

\begin{figure}[t]
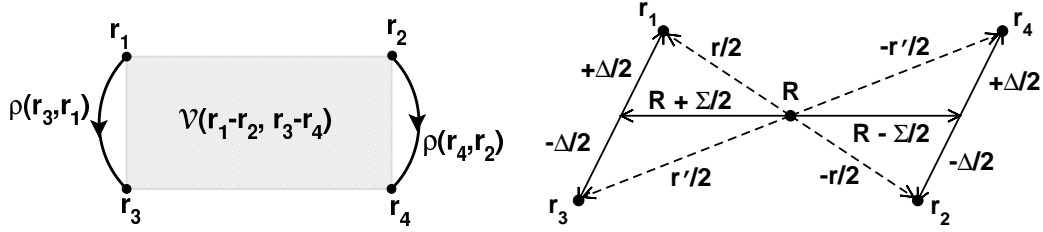

 \begin{center}
  \includegraphics*[width=2.6in]{fig_dme_Vblob}
  \hspace*{.1in}
  \includegraphics*[width=2.6in]{fig_dme_nonlocal2}
 \end{center}
 \caption{(a)~Schematic diagram for approximations
 to $\Wint$ that can be
 expanded using the DME. 
 (b)~Coordinates appropriate for the DME applied
 to the Hartree-Fock potential energy
 with a non-local potential.}
 \label{fig:dme_nonlocal}
\end{figure}

In this initial work we will only consider terms in the energy functional
arising from products of the scalar-isoscalar ($\rho_0$) density matrices in
Eq.~(\ref{eq:HF_DM1}), which 
are the relevant terms for spin-saturated systems with $N=Z$. 
Thus, we will drop the subscript ``$0$'' on the density matrices from now on.

After switching to relative/center-of-mass (COM) coordinates
(see Fig.~\ref{fig:dme_nonlocal}) and noting that the
free-space two-nucleon potential is diagonal in the COM coordinate, 
the starting 
point for our DME of the two-body Hartree-Fock contribution 
from a non-local interaction is 
\bea
\label{eq:hf_potential_contr}
\WHF &=&\frac{1}{32}
\int d{\bf R}\,d{\bf r}\,
d{\bf r'}\,\rho({\bf R}+\frac{\bf r'}{2},{\bf R}+\frac{\bf r}{2})
\rho({\bf R}-\frac{\bf r'}{2},{\bf R}-\frac{\bf r}{2}) 
 \tr_{\sigma\tau}[\langle\rvec|\antisymV|\rvec'\rangle] \;,
\nonumber
\\
\eea
where $\antisymV$ denotes the antisymmetrized interaction
and the trace is defined as
\beqn
\tr_{\sigma\tau}[\langle\rvec|\antisymV|\rvec'\rangle] \equiv
\sum_{\{\sigma\tau\}} \langle \rvec\sigma_1\tau_1\sigma_2\tau_2|
\widehat{V}(1-P_{12})|\rvec'\sigma_1\tau_1\sigma_2\tau_2\rangle
\;.
\eeqn
The DME derivation of Negele and
Vautherin (NV)~\cite{NEGELE72} 
focuses
on applications to local potentials,
which satisfy 
$\langle\rvec |\widehat{V}
|\rvec'\rangle = \delta(\rvec-\rvec')\langle
\rvec|\widehat{V}|\rvec'\rangle$. 
While the original NV work included 
coordinate-space formulas applicable for non-local interactions%
\footnote{However, note that the final formulas for non-local potentials in
Ref.~\cite{NEGELE72} have numerous errors, which were not among those
corrected in Ref.~\cite{NEGELE75}.}, for
low-momentum potentials it is convenient to revisit
and extend the original derivation to a momentum-space formulation.
We note that Kaiser et al.\ have shown how to
use medium-insertions in momentum space in their application of
the DME to chiral perturbation theory at finite 
density~\cite{Kaiser:2002jz,Kaiser:2003uh,Kaiser:2005}.

For the momentum space formulation, we first 
rewrite the density matrices appearing in
Eq.~(\ref{eq:hf_potential_contr}) as
\bea
  \rho({\bf R}\pm{\bf r'}/{2},{\bf R}\pm{\bf r}/{2})=
  \rho({\bf R}^\pm\pm{\bDel }/{2},{\bf R}^\pm\mp{\bDel}/{2})\;,
\eea
where the vectors appearing on the right-hand side are defined by  
(see Fig.~\ref{fig:dme_nonlocal})
\bea
  {\bf R}^\pm={\bf R}\pm\frac{1}{2}\bSig \;, \qquad
\bSig=\frac{1}{2}({\bf r'}+{\bf r})\;, \qquad
\Delta=\frac{1}{2}({\bf r'}-{\bf r}) \;.
\eea
Introducing the Fourier transform
of $\antisymV$ in the momentum transfers conjugate to
$\bSig$ and $\Delta$, 
\beqn
 \qvec=\kvec-\kpvec \;, \qquad \pvec = \kvec+\kpvec \;,
\eeqn
(where $\kpvec$, $\kvec$ correspond to 
relative momenta) gives
\bea
\label{eq:HF_kspace}
\WHF = \frac{1}{32}\int d\Rvec\int 
        \frac{d\qvec\,d\pvec}{(2\pi)^6}\,F(\Rvec,\qvec,\pvec)
   \, \tr_{\sigma\tau}[\widetilde{\antisymV}(\qvec,\pvec)]
  \;,
\eea
where we have defined
\bea
\label{eq:Fqp}
F(\Rvec,\qvec,\pvec)&\equiv&\int d\bSig\,d\bDel\, 
  e^{i\qvec\cdot\bSig}\,e^{i\pvec\cdot\bDel}
  \,\rho(\Rvec^+ -\bDel/2,\Rvec^+ +\bDel/2)
\\
&& \null \times
\rho(\Rvec^- +\bDel/2,\Rvec^- -\bDel/2)\;, \nonumber
\eea
and
\beqn
\label{eq:Vqp}
\widetilde{\antisymV}(\qvec,\pvec)
 \equiv 8\int d\bSig\,d\bDel\, e^{-i\qvec\cdot\bSig}\,
e^{-i\pvec\cdot\bDel}\,\langle\bSig-\bDel|\antisymV|\bSig+\bDel\rangle\;.
\eeqn
The momenta $\qvec$ and $\pvec$ correspond to the momentum transfers for a
local interaction in the direct and exchange channels. That is, the direct
matrix element is a function of $\qvec$ and the exchange is a function of
$\pvec$. In contrast, for a non-local interaction the direct and exchange matrix
elements depend on both   $\qvec$ and $\pvec$. This is the reason why we do not
attempt to separate out the Hartree (direct) and Fock (exchange) contributions
to $\WHF$, as is commonly done for local interactions.

The trace of Eq.~(\ref{eq:Vqp}) can be written in a more convenient form for
our purposes as a sum over partial wave matrix elements,
\bea
 \label{eq:Vqplsjt}
 \tr_{\sigma\tau}[\widetilde{\antisymV}(\qvec,\pvec)] =
 8\pi \sum_{lsj}\,^{'} (2j+1)(2t+1)\,P_{l}(\widehat{\kvec}
 \cdot\widehat{\kpvec})\langle klsjt|V|k'lsjt\rangle\;,
\eea
 where the primed summation 
 means that it is restricted to values where $l+s+t$ is odd,
 with $\kvec = \frac{1}{2}(\pvec+\qvec)$ and $\kpvec=\frac{1}{2}
 (\pvec-\qvec)$. For simplicity we have assumed a 
 charge-independent two-nucleon interaction, although charge-dependence
 can easily be included. 
 
 \subsection{Density Matrix Expansion}

The expression Eq.~(\ref{eq:HF_kspace}) for $\WHF$ 
is written in terms of off-diagonal density 
matrices constructed from the Kohn-Sham orbitals.
Consequently, the corresponding $\Gamma_{\rm HF}[\rho]$
is an \emph{implicit} functional of the density. The 
orbital-dependent  $\Gamma_{\rm HF}$ requires   
the use of the functional derivative chain rule
to evaluate  
$J_1(\Rvec) =\delta \Gamma_{\rm HF}[\rho]/\delta \rho(\Rvec)$
in the self-consistent determination of the Kohn-Sham
potential, which presents computational challenges
and would require substantial enhancements
to existing Skyrme HFB codes. 

Alternatively,
we can apply Negele and Vautherin's DME to $\WHF$, resulting in an expression
as in Eq.~(\ref{eq:eDME})
with \emph{explicit} dependence on the local quantities
$\rho(\Rvec)$,  $\tau(\Rvec)$, and $|\nabla \rho(\Rvec)|^2$,
\beqn
  \WHF = \int d\Rvec\,( A[\rho] + B[\rho]\tau +
    C[\rho](\nabla\rho)^2 + \cdots)
  \;.
  \label{eq:WHFdme}
\eeqn
The starting point of the DME is the 
formal identity~\cite{NEGELE72}
\bea
  \rho({\bf R}+{\bf s}/{2},{\bf R}-{\bf s}/{2})&=&
  \sum_a \phi^{*}({\bf R}+{\bf s}/{2})\phi({\bf R}-{\bf s}/{2})
  \nonumber\\
  &=&\bigl[e^{{\bf s}\bfcdot(\nabla_1-\nabla_2)/2}\sum_a
  \phi^{*}({\bf R}_1)\phi({\bf R}_2)\bigr]_{{\bf R}_1={\bf R}_2={\bf R}}\;,
\eea 
where $\nabla_1$ and $\nabla_2$ act on ${\bf R}_1$ and ${\bf R}_1$,
respectively, and the result is evaluated at 
${\bf R}_1={\bf R}_1={\bf R}$. 
We assume here that
time-reversed orbitals are filled pairwise, so that the linear term of the
exponential expansion vanishes. Hence, through second-order gradient terms the
angular integral of the density matrix squared is equivalent to the integral of
the square of the angle-averaged density matrix. In this way, the leading
off-diagonal behavior of the density matrices in $\WHF$ is captured by
simpler expressions. 

The angle-averaged density matrix takes the form
\bea
  \hat{\rho}({\bf R}+{\bf s}/{2},{\bf R}-{\bf s}/{2})&=&
  \frac{1}{2}\int\!\hbox{d}\cos\theta\, \exp\bigl[{\bf s}\bfcdot
  (\nabla_1-\nabla_2)/2\bigr]\rho({\bf R}_1,{\bf R}_2)
  \nonumber\\
  &=&
  \left.\frac{\sinh[\frac{1}{2} s  |\nabla_1-\nabla_2|]}
  {\frac{1}{2}s |\nabla_1-\nabla_2|}\rho({\bf R}_1,{\bf R}_2)
  \right|_{\Rvec_1 = \Rvec_2 = \Rvec}
  \;,
\eea
with $s \equiv |\bf s|$.
Using a 
Bessel-function expansion
(which is simply the usual plane-wave expansion with real arguments),
\beqn
\label{eq_bessel_expansion}
  \frac{1}{x y}\sinh(xy)=\frac{1}{x}\sum_{k=0}^\infty
  (-1)^k (4k+3)j_{2k+1}(x)\mathcal{Q}_k (y^2) \;,
\eeqn
where $\mathcal{Q}$ is related to the usual Legendre polynomial by
$\mathcal{Q}(z^2)={P_{2k+1}(iz)}/(iz)$, we can express the angle-averaged
density matrix as
\bea
\label{eq:canonicalDME}
  \hat{\rho}({\bf R}+{\bf s}/{2},{\bf R}-{\bf s}/{2})
   &=&
   \frac{1}{s \kf (\Rvec)}\Biggl[\sum_{n=0}^\infty(4n+3)j_{2n+1}(s \kf (\Rvec))
   \nonumber \\
   & & \qquad \null \times
  \mathcal{Q}_n\biggl(\biggl(\frac{\nabla_1-\nabla_2}{2\kf (\Rvec)}\biggr)^2\biggr)\Biggr]
  \rho({\bf R}_1,{\bf R}_2)\;,
\eea
where an arbitrary momentum scale $\kf (\Rvec)$ has been introduced. 
Equation~(\ref{eq:canonicalDME}) is 
independent of $\kf $ if all terms are kept, but any truncation will give
results depending on the particular choice for $\kf $. In this initial study, we
employ the standard LDA choice of Negele and Vautherin:
\beq
 \kf (\Rvec)=(3\pi^2 \rho({\bf R})/2)^{1/3}  \;. 
 \label{eq:standardLDA}
\eeq
Alternative  choices for $\kf (\Rvec)$ to optimize the
convergence of truncated expansions of Eq.~(\ref{eq:canonicalDME}) 
and to establish a power counting will be
explored in a future paper. 

Following Negele and Vautherin, Eq.~(\ref{eq:canonicalDME}) is 
truncated to terms with $n\leqslant 1$, which yields the fundamental 
equation of the DME,
\bea
\label{eq:DMEmastereqn}
\hat{\rho}({\bf R}+\frac{\bf s}{2},{\bf R}-\frac{\bf s}{2})
   & \approx & \rhoSL(\kf (\Rvec)s)\,\rho(\Rvec) 
\\
&&+ s^2 g(\kf (\Rvec)s)\bigl[\frac{1}{4}\nabla^2\rho(\Rvec) - \tau(\Rvec) 
+ \frac{3}{5}\kf (\Rvec)^2\rho(\Rvec)\bigr]\nonumber,
\eea
where
\beqn 
     \rhoSL(x) \equiv 3j_1(x)/x \;, \qquad 
           g(x) \equiv 35 j_3(x)/2x^3 \;,
\eeqn
and the  kinetic energy density is $\tau(\Rvec) = \sum_{i}|\nabla
\phi_i(\Rvec)|^2$. 
If a short-range interaction is folded with the density matrix,
then a truncated Taylor series expansion of Eq.~(\ref{eq:DMEmastereqn})
in powers of $s$
would be justified and would produce a quasi-local functional.
But the local $\kf $ in the interior of a nucleus
is typically greater than the pion mass $m_{\pi}$,
so such an expansion would give a poor representation of the
physics of the long-range pion exchange interaction.

Instead, the DME is constructed as an expansion about the exact
nuclear matter density matrix.
Thus,
Eq.~(\ref{eq:DMEmastereqn}) has the important feature
that it reduces to the density matrix in the homogenous nuclear matter
limit, $\rhoNM(\Rvec + {\bf s}/2, \Rvec - {\bf s}/2) = \rhoSL(\kf s)\,\rho$. 
As a result, the resummed expansion in
Eq.~(\ref{eq:DMEmastereqn}) 
 does not distort the finite range physics, as the
long-range one-pion-exchange 
contribution to nuclear matter is exactly reproduced and the
finite-range physics is encoded as non-trivial (e.g., non-monomial) density
dependence in the resulting functional. 
The small parameters justifying this expansion emerge in the functionals
as integrals over the inhomogeneities of the density.
(See Ref.~\cite{Bhattacharyya:2004qm} for examples
of estimated contributions to a functional for a model problem.)

In the case of a local interaction, 
the  Fock term is schematically given by $W_{\rm F} \sim \int
d\Rvec\,d{\bf s}\, \rho^2(\Rvec+\svec/2,\Rvec - \svec/2)V(\svec)$,
so a single application of
Eq.~(\ref{eq:DMEmastereqn}) is sufficient to cast $\WHF$ into the desired form.
For a
non-local interaction the calculation is more involved as two applications of
the DME are required. Following Negele and
Vautherin, we first rewrite the density matrices appearing in
Eq.~(\ref{eq:hf_potential_contr}) as
\bea
  \rho({\bf R}\pm{\bf r'}/{2},{\bf R}\pm{\bf r}/{2})=
  \rho({\bf R}^\pm\pm{\bDel }/{2},{\bf R}^\pm\mp{\bDel}/{2})\;,
\eea
where the vectors appearing on the right-hand side are defined by  
(see Fig.~\ref{fig:dme_nonlocal})
\bea
  {\bf R}^\pm\equiv{\bf R}\pm\frac{1}{2}\bSig \;, \qquad
\bSig\equiv\frac{1}{2}({\bf r'}+{\bf r})\;, \qquad
\Delta\equiv\frac{1}{2}({\bf r'}-{\bf r}) \;.
\eea
To simplify the notation we define
\bea
\kf ^{\pm}\equiv \kf ({\bf R}^{\pm}) \;,\quad \rho^{\pm}\equiv\rho({\bf
  R}^{\pm}) \;,\quad
\tau^\pm\equiv\tau({\bf R}^\pm) \;,
\eea
and it is from now on understood that the functions without superscripts
depend only on the center-of-mass vector ${\bf R}$ if the argument
is not written explicitly.

The first application of the DME corresponds to an expansion in the non-locality
$\bDel$ about the ``shifted'' COM coordinates $\Rvec^{\pm}$, giving
\bea
   \nonumber
   &&
   \rho({\bf R}+ \rvec'/2,{\bf R}+ \rvec/2)
   =
   \rho({\bf R}^+ - \Delta/2,{\bf R}^+ + \Delta/2)
   \\
   && \qquad\qquad \null \approx \rhoSL(\kf ^+\Delta)\rho^+
   +\Delta^2 g(\kf ^+\Delta)
   \bigl[\frac{1}{4}\nabla^2\rho^+-\tau^+
   +\frac{3}{5}{\kf ^+}^2\rho^+\bigr] \;.
\eea
Thus, we can expand the product of density matrices in
Eq.~(\ref{eq:hf_potential_contr}) as
\bea
  &&
  \rho({\bf R}+\frac{\bf r'}{2},{\bf R}+\frac{\bf r}{2})
 \rho({\bf R}-\frac{\bf r'}{2},{\bf R}-\frac{\bf r}{2}) 
   = 
  \rhoSL(\kf ^+\Delta)\rho^+\rhoSL(\kf ^-\Delta)\rho^-
   \nonumber \\
  && \qquad\qquad \null
  +\Delta^2 g(\kf ^+\Delta)\rhoSL( \kf ^-\Delta)\rho^-
        [\frac{1}{4}\nabla^2\rho^+-\tau^+
              +\frac{3}{5}{\kf ^+}^2\rho^+\bigr]
    \nonumber \\
  && \qquad\qquad \null
  +\Delta^2 g(\kf ^-\Delta)\rhoSL( \kf ^+\Delta)\rho^+
        [\frac{1}{4}\nabla^2\rho^--\tau^-
               +\frac{3}{5}{\kf ^-}^2\rho^-\bigr] \;,
\eea
where we have dropped terms quadratic in the gradient. 
We then define
\bea
  \alpha(\rho^\pm)=\rhoSL(\kf ^\pm\Delta)\rho^\pm \;,
\eea
and use Eq.~(\ref{eq_bessel_expansion}) to perform a \emph{second} 
density matrix expansion on $\alpha(\rho^{+})\alpha(\rho^{-})$ 
in $\bSig$ about $\Rvec$,
\beqn
  \alpha(\rho^+)\alpha(\rho^-) \approx \rhoSL(\kf \Sigma)\,
  \alpha^2
   +\frac{\Sigma^2}{2} g(\kf \Sigma)
   [\alpha\nabla^2 \alpha -|\nabla\alpha|^2+\frac{6}{5}\kf ^2\alpha^2] \;.
\eeqn

From a Taylor expansion of $\rhoSL(\kf \Sigma)$ and 
$g(\kf \Sigma)$ it is evident that the $(\kf \Sigma)^2$ coefficients
of $\alpha^2$ exactly cancel each other. Because we desire a final expression
that reproduces the exact nuclear matter limit (and the presence
of the $\rhoSL(\kf \Sigma)$ term spoils this limit), we follow the 
philosophy of Negele and Vautherin  and use this leading
cancellation to motivate a different rearrangement and truncation
of Eq.~(\ref{eq:canonicalDME}) such that
\bea
   \alpha(\rho^+)\alpha(\rho^-)&\approx&
   \alpha^2 +\frac{\Sigma^2}{2} g( \kf \Sigma)
   [\alpha\nabla^2 \alpha -|\nabla\alpha|^2] \;.
\eea
The freedom to rearrange the expansion as in the last
equation stems from the fact that the restriction of 
Eq.~(\ref{eq:canonicalDME})
to $n\leqslant 1$ 
terms gives a truncated expansion in powers of $\Sigma^2$. The
neglected terms, starting with $\Sigma^4$, involve higher derivatives of the
density. But having neglected these $\Sigma^4$ terms, 
retaining the other $\Sigma^4$ (and higher) contributions that are summed 
in $g(\kf \Sigma)$ is somewhat arbitrary. Therefore, Negele and Vautherin 
argue that it is 
advantageous to use this arbitrariness to ``reverse engineer'' the expansion
so that the exact nuclear matter limit is always exactly reproduced by
the leading term~\cite{NEGELE72}.
We emphasize that this is a prescription without established power
counting or error estimates, which must be assessed in future work.
As we show in Section~\ref{sec:results}, different prescriptions can lead
to significant changes in nuclear observables.

The gradient terms in the above equation can be evaluated with the
aid of the chain rule\footnote{Note that the equations here assume
the canonical choice of $\kf = (3\pi^2\rho/2)^{1/3}$. Alternative choices
for $\kf $, such as the one proposed by Campi and Bouyssy~\cite{Campi78} where
$\kf  = \kf (\rho,\nabla^2\rho,\tau)$ will generate additional terms by 
the chain rule.}
\beqn
  \nabla\alpha(\rho) = \nabla\rho
  \frac{\partial \alpha}{\partial \rho} \;,
  \qquad
  \nabla^2\alpha(\rho) = \nabla^2\rho\frac{\partial \alpha}{\partial \rho}
  +|\nabla\rho|^2\frac{\partial^2 \alpha}{\partial \rho^2} \;.
\eeqn
Recalling that we define the local Fermi momentum as
$\kf =(3\pi^2\rho)^{1/3}$, we can explicitly evaluate the first and second
derivatives of $\alpha$,
\beqn
\label{eq:del-alpha}
  \frac{\partial \alpha}{\partial \rho} = j_0(\kf \Delta)  \;, 
  \qquad
  \frac{\partial^2 \alpha}{\partial \rho^2} = 
  -\frac{\kf \Delta}{3\rho}j_1(\kf \Delta) \;.
\eeqn

Pulling it all together, the product of density matrices in 
Eq.~(\ref{eq:hf_potential_contr})
are approximately given in terms of local quantities by 
\bea
 &&
 \rho({\bf R}+\frac{\bf r'}{2},{\bf R}+\frac{\bf r}{2})
 \rho({\bf R}-\frac{\bf r'}{2},{\bf R}-\frac{\bf r}{2}) 
   \approx \rhoSL^2(\kf \Delta)\rho^2
    +\frac{1}{2} \Sigma^2 g(\kf \Sigma)\nonumber
  \\
  && \qquad\qquad\null
  \times \bigl(\rho\nabla^2\rho\, \rhoSL(\kf \Delta) j_0(\kf \Delta) 
    -|\nabla\rho|^2 
   [j_0^2(\kf \Delta)+j_1^2(\kf \Delta)]
  \bigr)
   \nonumber\\
  && \qquad\qquad\qquad \null + 2\Delta^2 g(\kf \Delta)\rhoSL(\kf \Delta)
  \bigl(\frac{1}{4} \rho\nabla^2\rho
     -\rho\,\tau  +\frac{3}{5}\kf ^2\rho^2
  \bigr) \;. 
  \label{eq:expandedDMs_non-local}
\eea

\subsection{Evaluation of $F(\Rvec, \qvec,\pvec)$ 
   and the DME coupling functions}

In the momentum space expression
for $\WHF$, it remains to 
evaluate the Fourier transforms defined in Eq.~(\ref{eq:Fqp})
for the expanded density matrices in Eq.~(\ref{eq:expandedDMs_non-local}). 
Identifying the terms in Eq.~(\ref{eq:WHFdme}) that give
the DME functionals $A[\rho]$, $B[\rho]$, and $C[\rho]$, we have
\bea
\label{eq:Fqp_A}
  F(\Rvec,\qvec,\pvec)\bigr |_{A} 
  &=& (2\pi)^3\delta(\qvec)\,\frac{4\pi}{\kf ^3}\,
  \bigl[ I_1(\bar{p}) + \frac{6}{5}I_2(\bar{p}) \bigr ]\, \rho^2 \;,
  \\
  \label{eq:Fqp_B}
  F(\Rvec,\qvec,\pvec)\bigr |_{B} 
  &=& -(2\pi)^3\delta(\qvec)\,\frac{8\pi}{\kf ^5}\,I_2(\bar{p})
  \rho\,\tau \;,
  \\
  F(\Rvec,\qvec,\pvec)\bigr |_{C} 
  &=& -\frac{8\pi^2}{\kf ^8}I_3(\bar{q})\,I_5(\bar{p})|\nabla \rho|^2
   +\bigl [ (2\pi)^3\delta(\qvec)\,\frac{2\pi}{\kf ^5}\,I_2(\bar{p})
  \nonumber
  \\
  \label{eq:Fqp_C}
  &&\qquad\quad 
  \null + \frac{8\pi^2}{\kf ^8}\, 
  I_3(\bar{q})\,I_4(\bar{p})\bigr ] \,\rho\,\nabla^2\rho\;,
\eea
where $\bar{p} = p/\kf $ etc., and the $\Rvec$-dependence of $\kf $, $\rho$, and
$\tau$ has been suppressed.  
The functions $I_{j}(\bar{p})$ and $I_j(\bar{q})$ are
simple polynomials (and theta functions)
in the scaled momenta $\bar{p}$  and $\bar{q}$: 
\bea
  \label{eq:DMEintegrals_NN1}
  I_1(\bar{p}) &\equiv& \int x^2\,dx\, j_0(\bar{p}x)\,\rhoSL^2(x) 
  = \frac{3\pi}{32} (16 - 12\pb + \pb^3)\,\theta(2-\pb) \;, \\
  I_2(\bar{p}) &\equiv& \int x^4\,dx\, j_0(\bar{p}x)\, \rhoSL(x) \,g(x) 
  \nonumber \\
  &&
  \qquad\qquad\qquad
  =-\frac{35\pi}{128} (\pb^5-18\pb^3 
     +40\pb^2-24\pb) \, \theta(2-\pb) \;,
  \label{eq:DMEintegrals_NN2}
  \\
  \label{eq:DMEintegrals_NN3}
  I_3(\bar{q}) &\equiv& \int x^4\, dx\, j_0(\bar{q}x) \, g(x) =
  -\frac{35\pi}{8} (5\qb^2-3) \, \theta(1-\qb) \;, \\
  \label{eq:DMEintegrals_NN4}
  I_4(\bar{p}) &\equiv& \int x^2 \,dx\, j_0(\bar{p}x)\, j_0(x)\, \rhoSL(x) =
  \frac{3\pi}{8} (2-\pb) \, \theta(2-\pb) \;, \\
  \label{eq:DMEintegrals_NN5}
  I_5(\bar{p}) &\equiv&\int x^2 \, dx\, j_0(\bar{px})
            [j_0^2(x) + j_1^2(x)]
  = \frac{\pi}{8\pb}  (4-\pb^2) \, \theta(2-\pb)\;.
\eea
Note that the trivial angular dependence of 
Eqs.~(\ref{eq:DMEintegrals_NN1})--(\ref{eq:DMEintegrals_NN5})
is a consequence of the angle averaging that is implicit with each application 
of the DME.  

With the aid of Eqs.~(\ref{eq:Fqp_A})--(\ref{eq:DMEintegrals_NN5}), 
we can now obtain
explicit expressions for the $A$, $B$, and $C$ coupling functions by grouping
terms appropriately and performing the relevant angular integrals. The
expressions for $A$ and $B$ follow immediately and are given by
\bea
  \label{eq:AB_NN}
  A[\rho] &=& \frac{\rho^2}{16\pi \kf ^3}\int_{0}^{2\kf }\!p^2dp\,  
  \tr_{\sigma\tau}
  [\widetilde{\antisymV}(0,\pvec)] 
     \, (I_1(\pb) + \frac{6}{5}I_2(\pb) ) \;,  \\
  B[\rho] &=& -\frac{\rho}{8\pi \kf ^5}\int_{0}^{2\kf }\!p^2dp\,  
  \tr_{\sigma\tau}
  [\widetilde{\antisymV}(0,\pvec)]\, I_2(\pb)\;,
\eea
where $\tr_{\sigma\tau}[\widetilde{\antisymV}(0,\pvec)]$ 
is given by a
simple sum of diagonal matrix elements in the different partial waves, 
\beqn
  \tr_{\sigma\tau}[\widetilde{\antisymV}(0,\pvec)] =
  8\pi  \sum_{lsj}\,^{'} (2j+1)(2t+1)\,\langle 
  \frac{p}{2}lsjt|V|\frac{p}{2}lsjt\rangle \;.
\eeqn
The primed sum is over all channels for which $l+s+t$ is odd.

The contributions to $\WHF$ that have gradients of the local density
take the form
\beq
  \WHF\bigr|_{|\nabla\rho|^2} = \int d\Rvec\,
    \bigl( C_{\nabla^2\rho}\nabla^2\rho(\Rvec)
  + C_{|\nabla\rho|^2}|\nabla\rho(\Rvec)|^2\bigr) 
   \;.
\eeq
We can perform a partial integration on the $\nabla^2\rho$ terms to cast
them into the canonical form proportional to only $|\nabla \rho|^2$;
that is, 
\beq
  \WHF\bigr|_{|\nabla\rho|^2} 
  =  \int d\Rvec\,|\nabla\rho(\Rvec)|^2\,\bigl[ C_{|\nabla\rho|^2} 
  -\frac{d}{d\rho}C_{\nabla^2\rho}\bigr] \;,
\eeq
so that
\beqn
  C[\rho] = C_{|\nabla\rho|^2} - \frac{d}{d\rho}C_{\nabla^2\rho}\;.
  \label{eq:deriv}
\eeqn
In practice it is efficient and accurate to calculate the derivative
in Eq.~\eqref{eq:deriv} numerically rather than analytically. 

The expressions for $C_{|\nabla\rho|^2}$ and $C_{\nabla^2\rho}$ are obtained by
substituting the relevant terms in $F(\Rvec,\qvec,\pvec)$ 
[see Eqs.~(\ref{eq:Fqp_A})--(\ref{eq:Fqp_B})] 
into Eq.~(\ref{eq:HF_kspace}) and performing
the angular integrals,
\bea
  C_{|\nabla\rho|^2} &=& 
    \frac{1}{32}\int\frac{d\qvec\, d\pvec}{(2\pi)^6}\, 
  \bigl( -\frac{8\pi^2}{\kf ^8}I_3(\qb)\, I_5(\pb) \bigr)  \,
    \tr_{\sigma\tau}
  [\widetilde{\antisymV}(\qvec,\pvec)]
  \\
  &=& -\frac{1}{16\pi^2\kf ^8}\int_0^{\kf}\! q^2dq \int_0^{2\kf}\! p^2dp\,  
  I_3(\qb)\, I_5(\pb) \,
  \widetilde{\antisymV}_{av}(q,p)
  \;,
\eea
\bea
  C_{\nabla^2\rho} &=& 
      \frac{\rho}{32}\int\frac{d\qvec\, d\pvec}{(2\pi)^6}\, 
    \bigl(\frac{1}{\kf ^5}(2\pi)^4 \delta^3(\qvec)\,I_2(\pb) 
    +\frac{8\pi^2}{\kf ^8}I_3(\qb)\, I_4(\pb)\bigr) \, 
      \tr_{\sigma\tau}
    [\widetilde{\antisymV}(\qvec,\pvec)]
  \nonumber\\
  &=& \frac{\rho}{32\pi \kf ^5}
    \int_0^{2\kf}\! p^2dp \,I_2(\pb)\, 
   \tr_{\sigma\tau}[\widetilde{\antisymV}(0,\pvec)]
  \nonumber\\
  &&\qquad\quad \null + 
    \frac{\rho}{16\pi^2 \kf ^8}\int_0^{\kf}\! q^2dq \int_0^{2\kf}\! p^2dp\,
    I_3(\qb)\,I_4(\pb)\,
   \widetilde{\antisymV}_{\rm av}(q,p)
   \;,
\eea
where $\widetilde{\antisymV}_{\rm av}(q,p)$ is the angle-averaged interaction,
\beqn
  \widetilde{\antisymV}_{av}(q,p) \equiv \frac{1}{2}\int d(\cos{\theta})\,
   \tr_{\sigma\tau} [\widetilde{\antisymV}(\qvec,\pvec)]  \;,
\eeqn
and $\widetilde{\antisymV}(\qvec,\pvec)$ is given 
by Eq.~(\ref{eq:Vqplsjt}). 
Note that care must be taken in the evaluation of  
$d C_{\nabla^2\rho}/d\rho$ if the vertex  
$\widetilde{\antisymV}(\qvec,\pvec)$ is density-dependent 
or if the local Fermi momentum is not taken to be
$\kf  = (3\pi^2\rho/2 )^{1/3}$. 

\section{DME for three-body potentials in momentum space}
\label{sec:dme-three-body}

In this section we extend the DME as applied to the
Hartree-Fock energy to include three-body
force contributions. The low-momentum 
interactions currently in use do not yet include consistently evolved 
three-body forces because of technical 
difficulties in carrying out the momentum-space evolution%
\footnote{However, the recent application of similarity 
renormalization group (SRG) methods to inter-nucleon 
potentials provide a computationally feasible path to 
the momentum-space evolution of many-body 
forces~\cite{Bogner:2006pc,Jurgenson2008}.}.
Therefore, as an approximation to the evolution, two 
short-distance low-energy constants in the leading 
chiral three-body force (this is N$^2$LO according to 
the power counting of Refs.~\cite{chiral3NF1,chiral3NF2}) 
are fit at each cutoff 
to properties of the triton and $^4$He to 
determine the three-body force. In the present work, 
we will use this force exclusively and postpone the 
treatment of general non-local three-body forces, as will be 
produced by an SRG evolution.

\subsection{$\WHF$ for local three-body forces}

The Hartree-Fock 3NF contribution to the total energy is given by
\beqn
  \WHF^{(3N)}=\frac{1}{6}\sum_{ijk}^{A}
  \langle i \,j \,k|V \mathcal{A}_{123}| i \,j \,k \rangle\;,
  \label{eq:HF3N}
\eeqn
where the summation is over the occupied Kohn-Sham states and
the operator $\mathcal{A}_{123}$ is the (un-normalized) three-nucleon
antisymmetrizer
\bea
  \mathcal{A}_{123}&=&(1+P_{13} P_{12}+P_{23} P_{12})(1-P_{12}) 
  \nonumber \\
  &=& (1 + P_{13}P_{23} + P_{12}P_{23})(1-P_{23} )
  \nonumber \\
  &=& (1+P_{23}P_{13} + P_{12}P_{13})(1-P_{13})\;.
\eea
Decomposing the three-body potential in the standard
fashion~\cite{fewbody},
\beqn
  V=V^{(1)}+V^{(2)}+V^{(3)} \;,
\eeqn
where $V^{(i)}$ is symmetric under $j\leftrightarrow k$,  we can write the full interaction
in terms of one component
\beqn
V = V^{(1)} + P_{23}P_{13}V^{(1)}P_{13}P_{23} 
  + P_{23}P_{12}V^{(1)}P_{12}P_{23} \;,
\eeqn
and so on. This allows us to simplify Eq.~(\ref{eq:HF3N}) by using
\beqn
 V\mathcal{A}_{123} = (1+P_{23}P_{13} + P_{23}P_{12})V^{(1)}\mathcal{A}_{123}
  \;, 
\eeqn
the cyclic nature of the trace along with $(1+P_{23}P_{12} + P_{23}P_{13})
\mathcal{A}_{123} = 3\mathcal{A}_{123}$,  
and other permutation operator identities to obtain
\bea
  \WHF^{(3N)} &=& \frac{1}{2}\sum_{ijk}^{A}\langle ijk|V^{(1)}
     \mathcal{A}_{123}|ijk\rangle \nonumber \\
   &=& \frac{1}{2}\sum_{ijk}^{A}\langle ijk|V^{(1)}(1+P_{23}P_{12}+
       P_{13}P_{12} - P_{12} - P_{23} - P_{13})|ijk\rangle 
       \nonumber \\
       &=&  \frac{1}{2}\sum_{ijk}^{A}\langle ijk|V^{(1)}(1+2P_{23}P_{12}
        - 2P_{12} - P_{23} )|ijk\rangle  \;.
  \label{eq:simplified3NHF}
\eea

Because the leading chiral EFT 3NF has a vanishing direct piece, there are only
three independent contributions to $\WHF$ that need to be evaluated: one
double-exchange term involving two permutation operators and two
single-exchange contributions. Writing Eq.~(\ref{eq:simplified3NHF})  
in terms of
density matrices and separating out the scalar-isoscalar contributions to
$\WHF^{(3N)}$ arising from single-exchange terms gives

\bea
\label{eq:HF1x}
W^{(1x)}_{\rm HF} &=& -\frac{1}{64}\int d\xone\, d\xtwo\, d\xthree\, \Bigl\{ 
                \nonumber \\
		&& \qquad\quad
		\rho(\xtwo,\xone) \rho(\xone,\xtwo) \rho(\xthree)
                \, \tr_{123} [ V^{(1)}(\xone,\xtwo,\xthree) 
                P^{\sigma\tau}_{12} ]
                \nonumber \\
        && \qquad \null +   \frac{1}{2} \rho(\xthree,\xtwo) \rho(\xtwo,\xthree) \rho(\xone)
                \, \tr_{123} [ V^{(1)}(\xone,\xtwo,\xthree) 
                P^{\sigma\tau}_{23} ] \Bigr\}
                \nonumber \\
         &=& -\frac{1}{64}\int d\xone\, d\xtwo\, d\xthree\, \Bigl \{ 
                \rho(\xtwo,\xone) \rho(\xone,\xtwo) \rho(\xthree)\,
	  \nonumber \\	
	   && \qquad \null \times  \Bigl (
                \int \frac{d\qvec_2\, d\qvec_3}{(2\pi)^6} 
		e^{-i\qvec_2\cdot(\xvec_{1}-\xvec_2)}
                e^{-i\qvec_3\cdot(\xvec_{1}-\xvec_3)} 
          \tr_{123} [\widetilde{V}^{(1)}(\qvec_2,\qvec_3)
	       P^{\sigma\tau}_{12}]\Bigr )
	    \nonumber \\   
         && \qquad\null +  
	   \frac{1}{2} \rho(\xthree,\xtwo) \rho(\xtwo,\xthree) \rho(\xone)
         \nonumber \\
         &&\qquad\null\times\Bigl(  \int \frac{d\qvec_2\, d\qvec_3}{(2\pi)^6}
                e^{-i\qvec_2\cdot(\xvec_{1}-\xvec_2)}
                e^{-i\qvec_3\cdot(\xvec_{1}-\xvec_3)} 
          \tr_{123}[\widetilde{V}^{(1)}(\qvec_2,\qvec_3)P^{\sigma\tau}_{23}]\Bigr )
           \Bigr \} \;,
\nonumber
\\
\eea
where $\tr_{123}\equiv \tr_{\sigma_1\tau_1} \tr_{\sigma_2\tau_2}
\tr_{\sigma_3\tau_3}$  and a local 3NF has been assumed. Similarly, the
scalar-isoscalar contributions to $\WHF$ arising from the
double-exchanges are given by
\bea
\label{eq:HF2x}
W^{(2x)}_{\rm HF} &=& \frac{1}{64}\int d\xone\, d\xtwo\, d\xthree\, 
      \rho(\xone,\xtwo) \rho(\xtwo,\xthree) \rho(\xthree,\xone)\,
       \Bigl\{
      \nonumber \\
      &&
      \int \frac{d\qvec_2\, d\qvec_3}{(2\pi)^6} e^{-i\qvec_2\cdot\xvec_{12}}
      e^{-i\qvec_3\cdot\xvec_{13}} 
     \tr_{123} [\widetilde{V}^{(1)}(\qvec_2,\qvec_3)
     P^{\sigma\tau}_{23}P^{\sigma\tau}_{12}] \Bigr\}
     \;,
\eea
where the Fourier transformed 3NF components are defined by
\beqn
  \langle \kvec_1\kvec_2\kvec_3|V^{(1)}|\kvec'_1\kvec'_2 \kvec'_3\rangle = 
  \bigl(\frac{2\pi}{\Omega}\bigr)^3\delta(\qvec_1+\qvec_2+\qvec_3) 
  \widetilde{V}^{(1)}(\qvec_2,\qvec_3)
  \;.
\eeqn
Here $\Omega$ is the volume (which drops out of all final expressions) and 
$\qvec_i = \kvec_i - \kvec'_i$ is the momentum transfer.

\begin{figure}[tp]
 \begin{center}
  \includegraphics*[width=4.0in]{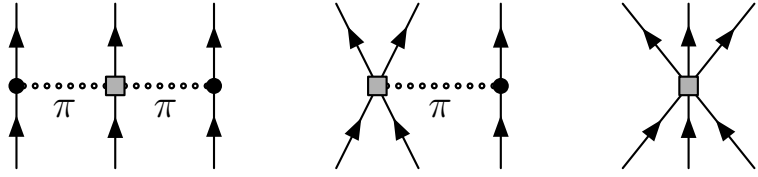}
 \end{center}
 \vspace*{-.1in}
 \caption{The chiral three-body force at N$^2$LO according to
 the power counting of Ref.~\cite{N3LOEGM},
 which has a long-range $2\pi$-exchange part $V_c$ (left),
 an intermediate-range $1\pi$-exchange part $V_D$ (middle), and
 a short-range contact interaction $V_E$ (right).} 
 \label{fig:3bodypion}
\end{figure}

As discussed above, we approximate the RG evolution of the 
3N force with the leading-order chiral 3N force, which is comprised of
a long-range $2 \pi$-exchange part $V_c$, an intermediate-range $1 \pi$-exchange 
part $V_D$ and a short-range contact interaction 
$V_E$~\cite{chiral3NF1,chiral3NF2}, see Fig.~\ref{fig:3bodypion}. 
The $2 \pi$-exchange interaction is 
\beqn
  \widetilde{V}_c^{(k)}(\qvec_i,\qvec_j) = \biggl( \frac{g_A}{2 f_\pi} \biggr)^2 
  \frac{({\bm \sigma}_i \cdot {\bf q}_i) ({\bm \sigma}_j \cdot 
  {\bf q}_j)}{(q_i^2 + m_\pi^2) (q_j^2 + m_\pi^2)} \: F_{ijk}^{\alpha\beta} \,
  \tau_i^\alpha \, \tau_j^\beta \;,
  \label{eq:Vc}
\eeqn
where $F_{ijk}^{\alpha\beta}$ is defined as
\beqn
  F_{ijk}^{\alpha\beta} = \delta^{\alpha\beta} \Bigl[ - \frac{4 c_1 
  m_\pi^2}{f_\pi^2} + \frac{2 c_3}{f_\pi^2} \: {\bf q}_i \cdot {\bf q}_j
  \Bigr] + \sum_\gamma \, \frac{c_4}{f_\pi^2} \: \epsilon^{\alpha\beta\gamma}
  \: \tau_k^\gamma \: {\bm \sigma}_k \cdot ( {\bf q}_i \times {\bf q}_j )
   \;,
\eeqn
while the $1 \pi$-exchange and contact interactions are, respectively,
\begin{align}
\widetilde{V}_D^{(k)}(\qvec_i,\qvec_j) &= - \frac{g_A}{4 f_\pi^2} \, \frac{c_D}{f_\pi^2 \la_\chi}
\: \frac{{\bm \sigma}_j \cdot {\bf q}_j}{q_j^2 + m_\pi^2} \: ({\bm \tau}_i
\cdot {\bm \tau}_j) \, ({\bm \sigma}_i \cdot {\bf q}_j) \;, \\
\widetilde{V}_E^{(k)}(\qvec_i,\qvec_j) &= \frac{c_E}{ f_\pi^4 \la_\chi} \: ({\bm \tau}_i
\cdot {\bm \tau}_j) \;.
  \label{eq:VDVE}
\end{align}
In applying Eqs.~(\ref{eq:Vc})--(\ref{eq:VDVE}), 
we use $g_A = 1.29$, $f_\pi = 92.4 \mev$ and $m_\pi = 138.04 \mev$
and the $c_i$ constants extracted by the Nijmegen group in a partial
wave analysis with chiral $2 \pi$-exchange~\cite{const}. These are
$c_1 = -0.76 \gevi$, $c_3 = -4.78 \gevi$ and $c_4 = 3.96 \gevi$. 
Fit values for the $c_D$ and $c_E$ low-energy constants 
consistent with a sharply cutoff low-momentum potential
are tabulated in Ref.~\cite{Vlowk3NF} for $\la_\chi = 700 \mev$.
 
From the previous general expressions for $W^{1x}_{\rm HF}$ and
$W^{2x}_{\rm HF}$, we  need to evaluate the spin-isospin traces
$\tr_{123}[\widetilde{V}^{(1)}P^{\sigma\tau}_{12}]$, 
$\tr_{123}[\widetilde{V}^{(1)}P^{\sigma\tau}_{23}]$,
and $\tr_{123}[\widetilde{V}^{(1)}P^{\sigma\tau}_{23}P^{\sigma\tau}_{12}]$. 
For the single-exchanges  we find
\bea
  \tr_{123} [\widetilde{V}_E^{(1)}(\qvec_2,\qvec_3)P^{\sigma\tau}_{23}] &=&
  48\, \frac{c_E}{ f_\pi^4 \la_\chi}  \;, 
  \\
\tr_{123}[\widetilde{V}_D^{(1)}(\qvec_2,\qvec_3)P^{\sigma\tau}_{23}] &=&
 -48 \,\frac{g_A}{4 f_\pi^2} \, 
 \frac{c_D}{f_\pi^2 \la_\chi}\frac{q_3^2}{q_3^2 + m_\pi^2} \;, 
 \\
 \tr_{123}[\widetilde{V}_c^{(1)}(\qvec_2,\qvec_3)P^{\sigma\tau}_{23}] &=&
 48\, \bigl( \frac{g_A}{2 f_\pi} \bigr)^2 \frac{\qvec_2\cdot\qvec_3}
 {(q_2^2 + m_\pi^2)(q_3^2 + m_\pi^2)} 
  \nonumber \\ 
  &&\qquad\qquad \null \times\, \bigl[ - \frac{4 c_1 
  m_\pi^2}{f_\pi^2} + \frac{2 c_3}{f_\pi^2} \: {\bf q}_2 \cdot {\bf q}_3 \bigr]
 \;, \\
 \tr_{123}[\widetilde{V}_E^{(1)}(\qvec_2,\qvec_3)P^{\sigma\tau}_{12}] &=& 
 \tr_{123}[\widetilde{V}_D^{(1)}(\qvec_2,\qvec_3)P^{\sigma\tau}_{12}] \nonumber
 \\ &=&
  \tr_{123}[\widetilde{V}_c^{(1)}(\qvec_2,\qvec_3)P^{\sigma\tau}_{12}] = 0\;,
\eea
while the various double-exchange terms give
\bea
  \tr_{123}[\widetilde{V}_E^{(1)}(\qvec_2,\qvec_3)P^{\sigma\tau}_{23}
  P^{\sigma\tau}_{12}] &=&
  12\, \frac{c_E}{ f_\pi^4 \la_\chi} \;, 
  \\ 
  \tr_{123}[\widetilde{V}_D^{(1)}(\qvec_2,\qvec_3)
      P^{\sigma\tau}_{23}P^{\sigma\tau}_{12}]
  &=& -12   \frac{g_A}{4 f_\pi^2} \, 
    \frac{c_D}{f_\pi^2 \la_\chi}\frac{q_3^2}{q_3^2 + m_\pi^2} \;,
  \\
  \tr_{123}[\widetilde{V}_c^{(1)}(\qvec_2,\qvec_3)P^{\sigma\tau}_{23}P^{\sigma\tau}_{12}]
  &=& 12\, \bigl( \frac{g_A}{2 f_\pi} \bigr)^2 \frac{\qvec_2\cdot\qvec_3}
  {(q_2^2 + m_\pi^2)(q_3^2 + m_\pi^2)}
  \nonumber \\ 
  &&\quad \times\, \bigl[ - \frac{4 c_1 
  m_\pi^2}{f_\pi^2} + \frac{2 c_3}{f_\pi^2}\bigl(1 + c_4/c_3\bigr) 
    \: {\bf q}_2 \cdot {\bf q}_3 \bigr]
  \nonumber \\
  &&\quad - 24 \,\bigl( \frac{g_A}{2 f_\pi} \bigr)^2\frac{c_4}{f_\pi^2}
       \frac{q_2^2q_3^2}{(q_2^2 + m_\pi^2)(q_3^2 + m_\pi^2)} \;.
\eea

Note that for the $V_E$ and $V_D$ terms, it is not necessary to treat
separately the single- and double-exchange contributions because their structure
is identical due to the nature of the zero-range three- and two-body vertices.
Substituting the  spin-isospin-traced interactions into 
Eqs.~(\ref{eq:HF1x})--(\ref{eq:HF2x}) and simplifying gives
\bea
  \label{eq:W_ED}
  \WHF^{E} &=& -\frac{3}{16} g_E\int d\xvec \,[\rho(\xvec)]^3 \;, \\
  \WHF^{D} &=& \frac{3}{16} g_D
  \int d\xvec_2\, d\xvec_3\, [\rho(\xvec_2,\xvec_3)]^2\rho(\xvec_2)
  \\
  &&\qquad\qquad
  \null\times
  \Bigl(\int \frac{d\qvec_3}{(2\pi)^3}
     e^{-i\qvec_3\cdot(\xvec_2-\xvec_3)}\frac{q_3^2}{q_3^2 + m_\pi^2}\Bigr) 
     \nonumber  \;,
\eea
where $g_E\equiv  {c_E}/{ f_\pi^4 \la_\chi}$ and
$g_D\equiv ({g_A}/{4 f_\pi^2}) \, ({c_D}/{f_\pi^2 \la_\chi})$.
Similarly, the single- and double-exchange contributions from the
$2\pi$-exchange 3NF are given by
\bea
\label{eq:W_c1x}
W^{(1x,c)}_{\rm HF} &=& -\frac{3}{8}g_c \int d\xvec_1\, d\xvec_2\, d\xvec_3\, 
         \rho(\xvec_1) \rho(\xvec_2,\xvec_3) \rho(\xvec_3, \xvec_2)
	\nonumber \\
	&& \qquad \null\times \Bigl\{ 
      \int \frac{d\qvec_2\, d\qvec_3}{(2\pi)^6} \,
      e^{-i\qvec_2\cdot(\xvec_1 - \xvec_2)}
     e^{-i\qvec_3\cdot(\xvec_1-\xvec_3)} 
       \frac{\qvec_2\cdot\qvec_3}
         {(q_2^2 + m_\pi^2)(q_3^2 + m_\pi^2)} 
         \nonumber \\
       &&\qquad\qquad\times\,  \bigl[ - \frac{4 c_1 m_\pi^2}{f_\pi^2} + 
         \frac{2 c_3}{f_\pi^2} \: {\bf q}_2 \cdot {\bf q}_3\bigr] \Bigr\}\;,
 \eea
and
\bea
  \label{eq:W_c2x}
   W^{(2x,c)}_{\rm HF} &=& 
     \frac{3}{16}g_c \int d\xvec_1\, d\xvec_2\, d\xvec_3\, 
         \rho(\xvec_1,\xvec_2)
         \rho(\xvec_2,\xvec_3) \rho(\xvec_3, \xvec_1)
     \nonumber \\
    &&\qquad \null \times  \int \frac{d\qvec_2\, d\qvec_3}{(2\pi)^6}
     \, e^{-i\qvec_2\cdot(\xvec_1 - \xvec_2)}
          e^{-i\qvec_3\cdot(\xvec_1-\xvec_3)} 
      \Bigl\{ \frac{\qvec_2\cdot\qvec_3}
         {(q_2^2 + m_\pi^2)(q_3^2 + m_\pi^2)} 
         \nonumber \\
       &&\qquad\qquad\times\,  \bigl[ - \frac{4 c_1 m_\pi^2}{f_\pi^2} + 
         \frac{2 c_3}{f_\pi^2} (1+c_4/c_3 ) \: 
	   {\bf q}_2 \cdot {\bf q}_3\bigr]
         \nonumber \\
         &&\qquad\qquad\qquad - \frac{2c_4}{f_\pi^2}
	    \frac{q_2^2q_3^2}{(q_2^2 + m_\pi^2)
         (q_3^2 + m_\pi^2)}
          \Bigr \}\;,
\eea
where $g_c\equiv ({g_A}/{2 f_\pi})^2$.

\subsection{D-term}

As with the nucleon-nucleon contributions to $\WHF$, it is  convenient to
recast the 3NF Hartree-Fock expressions into momentum space. Changing to
relative/center-of-mass coordinates ($\Rvec = (\xvec_2+\xvec_3)/2$, $\rvec =
\xvec_2-\xvec_3$), the $1\pi$-exchange 3N Hartree-Fock contribution becomes 
\bea
   \WHF^{D} &=& \frac{3}{16} g_D
   \int d\Rvec\, d\rvec\, [\rho(\Rvec+\rvec/2,\Rvec -\rvec/2)]^2
   \rho(\Rvec + \rvec/2)
   \nonumber \\
   && \qquad\null\times
   \int \frac{d\qvec}{(2\pi)^3} \, e^{-i\qvec\cdot\rvec}
     \frac{q^2}{q^2 + m_\pi^2}
   \nonumber\\
   \label{eq:WD_qspace}
   &=& \frac{3}{16}g_D \int d\Rvec \int\frac{d \qvec}{(2\pi)^3}
   \, F(\Rvec,\qvec) \, \frac{q^2}{q^2 + m_\pi^2}  \;,
\eea
where we have defined
\beqn
  \label{eq:Fq_D}
  F(\Rvec,\qvec) \equiv \int d\rvec\, e^{-i\qvec\cdot\rvec}
  [\rho(\Rvec+\rvec/2,\Rvec -\rvec/2)]^2
  \rho(\Rvec + \rvec/2)  \;.
\eeqn

Applying the DME separately to the product of non-local 
and local densities in $F(\Rvec,\qvec)$ yields 
\bea
\nonumber
  && [\rho(\Rvec+\rvec/2,\Rvec-\rvec/2)]^2  \approx
      \rhoSL(\kf r) \rho 
    + r^2g(\kf r)\bigl[ \frac{1}{2}\rho\nabla^2\rho
     - 2\rho\tau  +\frac{3}{5}\kf ^2\rho^2\bigr]
     \;,\\
\eea
and
\bea
  \rho(\Rvec + \rvec/2) \approx &&\rhoSL(\kf r)\rho 
    + r^2g(\kf r)\bigl[ \frac{1}{4}\nabla^2\rho 
  + \frac{3}{5}\kf ^2 \rho \bigr]\;.
\eea

Combining the two expansions and dropping terms of higher order 
in the DME, we find
\bea
  && [\rho(\Rvec+\rvec/2,\Rvec-\rvec/2)]^2\rho(\Rvec + \rvec/2) \approx
  \rhoSL^2(\kf r)\rho^3 
  \nonumber \\
  && \qquad\qquad \null + r^2g(\kf r)\rhoSL(\kf r)\Bigl[
  \frac{3}{4}\rho^2\nabla^2\rho - 2\rho^2\tau + \frac{6}{5}\kf ^2\rho^3\Bigr]\;,
\eea 
where the $\Rvec$-dependence of $\kf $ and the local densities has been 
suppressed. Evaluating the Fourier transform defined in
Eq.~(\ref{eq:Fq_D})
using the approximate DME expressions and grouping terms according to which 
coupling function contribute gives
\bea
   F(\Rvec,\qvec)\bigr|_A &=& 4\pi \bigl(\frac{\rho}{\kf }\Bigr)^3 
                	 \bigl[I_1(\bar{q}) + \frac{6}{5}I_2(\bar{q})\bigr]
   \;, \label{eq:FA} \\
   F(\Rvec,\qvec)\bigr|_B &=& -\frac{8\pi \rho^2\tau}{\kf ^5}\,I_2(\bar{q})
   \;, \label{eq:FB} \\
   F(\Rvec,\qvec)\bigr|_C &=& \frac{3\pi}{\kf ^5}\,
        \rho^2\nabla^2\rho\,I_2(\bar{q}) \;,
   \label{eq:FC}
\eea
where the integrals $I_1(\bar{q})$ and $I_2(\bar{q})$
were defined in 
Eqs.~(\ref{eq:DMEintegrals_NN1})--(\ref{eq:DMEintegrals_NN2}) 
and $\bar{q} \equiv q/\kf $. Together with Eq.~(\ref{eq:WD_qspace}), we obtain 
the $1\pi$-exchange 3NF contributions to the EDF coupling functions

\bea
  A_{D}[\rho] &=& \frac{3\rho^3}{8\pi \kf ^3}g_D
  \int dq \,\frac{q^4}{q^2 + m_\pi^2} \,  
      [ I_1(\bar{q}) + \frac{6}{5}I_2(\bar{q})]  \;,
  \\
  B_{D}[\rho] &=& -\frac{3\rho^2}{4\pi \kf ^5}g_D
  \int dq \,\frac{q^4}{q^2 + m_\pi^2}\, 
      I_2(\bar{q}) \;,
  \\
  C_{D}[\rho] &=& -\frac{9}{32\pi}g_D \frac{d}{d\rho}
  \Bigl( \frac{\rho^2}{\kf ^5}\int dq 
    \frac{q^4}{q^2+m_\pi^2}I_2(\bar{q})\Bigr )\;.
\eea
 
 \subsection{c-term single-exchange}

Starting from the single-exchange HF contribution of the $2\pi$-exchange
3NF in Eq.~(\ref{eq:W_c1x}), we first change to Jacobi
coordinates,
\beqn
 \rvec_{23} =  \xvec_2 - \xvec_3 \;,
 \quad
 \rvec_{1} = \xvec_2 - \frac{1}{2}\bigl(\xvec_3+\xvec_1\bigr) \;,
 \quad 
 \Rvec = \frac{1}{3}\bigl( \xvec_1 + \xvec_2 + \xvec_3\bigr)\;,
\eeqn
followed by the change of momentum variables $\qvec \equiv \frac{1}{2}(\qvec_2
- \qvec_3)$ and  $\pvec = \qvec_2+\qvec_3$. The result is 
\bea
  \label{eq:W_c1x_jac}
  \WHF^{(1x,c)} &=&  
  -\frac{3}{8}g_c \int d\Rvec 
     \int\frac{d\qvec\, d\pvec}{(2\pi)^6} \,
       F_{1x}(\Rvec, \pvec,\qvec) \, V_{c_1c_3}(\pvec,\qvec)\;,
\eea
where $F_{1x}(\Rvec, \pvec, \qvec)$ is the Fourier transform of the product of
density matrices, 
\bea
  \label{eq:Fqp_1x}
  F_{1x}(\Rvec,\pvec,\qvec) 
    &=& \int d\rvec_1\, d\rvec_{23}\, e^{-i\pvec\cdot\rvec_1}\, 
             e^{i\qvec\cdot\rvec_{23}}\, \rho(\Rvec + 2\rvec_1/3)
             \\
    &&  \quad \null \times     
            [\rho(\Rvec -\rvec_1/3 +\rvec_{23}/2,
               \Rvec -\rvec_1/3 -\rvec_{23}/2)]^2 \nonumber\;,
\eea 
and $V_{c_1c_3}(\pvec,\qvec)$ is defined as
\bea
  V_{c_1c_3}(\pvec,\qvec) = \frac{\qvec_2\cdot\qvec_3}{(\qvec_2^2+m_\pi^2)
        (\qvec_3^2 + m_\pi^2)}\,\bigl[-\frac{4c_1m_\pi^2}{f_\pi^2}
          + \frac{2c_3}{f_\pi^2}\qvec_2\cdot\qvec_3\bigr]\;,
\eea
with $\qvec_2 = \pvec/2 + \qvec$ and $\qvec_3 = \pvec/2 -
\qvec$. 

Referring to Eq.~(\ref{eq:Fqp_1x}),  
we first expand $\rho(\xvec_2,\xvec_3)$ as
\bea
 \rho(\xvec_2,\xvec_3) &=& 
    \rho(\Rvec -\rvec_1/3 +\rvec_{23}/2,
         \Rvec -\rvec_1/3 -\rvec_{23}/2)
    \nonumber\\
   &\approx& \rhoSL(\kf (\Rvec^-)r_{23}) \, \rho(\Rvec^-) 
    + r_{23}^2 g(\kf (\Rvec^-)r_{23})
    \nonumber \\
   && \null \times 
       \bigl[ \frac{1}{4}\nabla^2\rho(\Rvec^-)
      - \tau(\Rvec^-)  + \frac{3}{5}\kf ^2(\Rvec^-)\rho(\Rvec^-)\bigr]
    \;, 
 \eea
where $\Rvec^- \equiv \Rvec - \rvec_1/3$. 
Performing a subsequent expansion about $\Rvec$ gives
\bea
\label{eq:dme_single_matrix}
 \rho(\xvec_2,\xvec_3) 
    &\approx & \rhoSL(\kf r_{23}) \, \rho 
    + r_{23}^2 g(\kf r_{23})
     \bigl[ \frac{1}{4}\nabla^2\rho
    - \tau  + \frac{3}{5}\kf ^2\rho\bigr] 
 \nonumber\\
    &&\hspace{40mm} + \frac{1}{9}r_1^2g(\kf r_{23}) \nabla^2
   \bigl(\rhoSL(\kf r_{23})\rho\bigr)\, ,
 \eea
where the second application of the DME has been modified slightly 
to ensure the leading term is exact in the nuclear matter limit. 
Similarly, the diagonal density $\rho(\xvec_1)$ is expanded as
\bea
\rho(\Rvec + \frac{2}{3}\rvec_1) \approx \rho 
   + \frac{4}{9}r_1^2g(\kf r_1)\nabla^2\rho\;.
\eea

Therefore, to second order in the DME we obtain
\bea
  && \rho(\Rvec + 2\rvec_1/3) [\rho(\Rvec -\rvec_1/3 +\rvec_{23}/2,
  \Rvec -\rvec_1/3 -\rvec_{23}/2)]^2 \approx  
   [\rhoSL(\kf r_{23})]^2\rho^3 
  \nonumber \\
  && \qquad                    
   \null + 2r_{23}^2g(\kf r_{23})\rhoSL(\kf r_{23})
  \bigl[ \frac{1}{4}\rho^2\nabla^2\rho -\rho^2 \tau 
  + \frac{3}{5}\kf ^2\rho^3\bigr]    
  \nonumber \\
  && \qquad\null
   + \frac{1}{9}r_1^2 g(\kf r_1)\nabla^2(\rhoSL(\kf r_{23})\rho)   
   + \frac{4}{9}r_1^2g(\kf r_1)\rhoSL^2(\kf r_{23})\rho^2\nabla^2\rho    
    \;.                                                                                                                                                  
  \label{eq:DME1x}
\eea
For the usual LDA choice for $\kf (\Rvec)$, 
the $\nabla^2(\rhoSL\rho)$ term evaluates to
\bea
 \nabla^2 (\rhoSL(\kf r_{23}) ) 
 &=& \Bigl[ \bigl\{\rhoSL(\kf r_{23}) 
 + \rho\frac{\partial}{\partial\rho}\rhoSL(\kf r_{23})\bigr\} 
 \nabla^2\rho \nonumber \\
 &&\quad + \bigl\{2\frac{\partial}{\partial\rho}\rhoSL(\kf r_{23}) +
  \rho\frac{\partial^2}{\partial\rho^2}\rhoSL(\kf r_{23})\bigr\} 
  |\nabla\rho|^2 \Bigr] \;,
\eea
which suggests a grouping of terms in Eq.~(\ref{eq:DME1x}) 
according to which coupling function they contribute to,
\bea
  \rho\cdot\rho^2\bigr|_A &=& \rhoSL^2(\kf r_{23})\rho^3 
    +\frac{6}{5}r_{23}^2g(\kf r_{23})\rhoSL(\kf r_{23})\kf ^2\rho^3 \;,  \\
  \rho\cdot\rho^2\bigr|_B &=& 
     -2r_{23}^2g(\kf r_{23})\rhoSL(\kf r_{23})\rho^2\tau \;, \\ 
  \rho\cdot\rho^2\bigr|_C &=& 
    \Bigl[ \frac{1}{2}r_{23}^2g(\kf r_{23})\rhoSL(\kf r_{23})\rho^2
  +\frac{2}{9}r_1^2g(\kf r_1)\rhoSL(\kf r_{23})\rho^2\ \nonumber \\
  &&\qquad \null \times \bigl\{3\rhoSL(\kf r_{23}) 
    +\rho\frac{\partial}{\partial\rho}\rhoSL(\kf r_{23})\bigr\}\Bigr] \
       \nabla^2\rho \nonumber \\
  && \qquad \null 
     +\Bigl[\frac{2}{9}r_1^2g(\kf r_1)\rhoSL(\kf r_{23})\rho^2
     \nonumber \\
  && \qquad \null \times   
     \bigl\{
     2\frac{\partial}{\partial\rho}\rho_{S}(\kf r_{23}) 
      + \rho\frac{\partial^2}{\partial\rho^2}\rhoSL(\kf r_{23})\bigr\}\Bigr]
      \bigl(\nabla\rho\bigr)^2 
      \;.
\eea

Evaluating the Fourier transform in Eq.~(\ref{eq:Fqp_1x}) gives
\bea
  \label{eq:F1x_A}
  F_{1x}(\Rvec,\pvec,\qvec)\bigr|_A 
  &=& (2\pi)^4\delta(\pvec)\frac{2\rho^3}{\kf ^3}
  \bigl[ I_1(\bar{q}) + \frac{6}{5}I_2(\bar{q})\bigr] \;,  \\
  \label{eq:F1x_B}
  F_{1x}(\Rvec,\pvec,\qvec)\biggr|_B &=& 
    -(2\pi)^4\delta(\pvec)\frac{4\rho^2\tau}{\kf ^5}\,I_2(\bar{q}) \;, \\
  F_{1x}(\Rvec,\pvec,\qvec)\biggr|_C &=& 
    \Bigl[(2\pi)^4\delta(\pvec)\frac{\rho^2}{\kf ^5}I_2(\bar{q})
    \nonumber\\
   &&  \null
    + \frac{32\pi^2\rho^2}{3\kf ^8}I_3(\bar{p})
     \bigl( I_1(\bar{q}) - \frac{1}{3}I_6(\bar{q})\bigr) \Bigr] 
     \nabla^2\rho \nonumber \\ 
   && \null - \Bigl[\frac{32\pi^2\rho}{9\kf ^8}I_3(\bar{p})
      \bigl\{I_6(\bar{q}) + \frac{2}{15}I_7(\bar{q}) 
      - \frac{1}{5} I_8(\bar{q})\bigr\}
      \Bigr]\bigl(\nabla \rho\bigr)^2 \;,
\label{eq:F1x_C}
\eea
where $I_1$--$I_5$ have been defined in 
Eqs.~(\ref{eq:DMEintegrals_NN1})--(\ref{eq:DMEintegrals_NN5}) 
and the new integrals $I_6$--$I_8$ are defined as
\bea
   I_6(\pb) &\equiv& \int x^2dx\, j_0(\pb x) \rhoSL(x)j_2(x) 
      = \frac{3\pi}{32}(8-8\pb + \pb^3)\,\theta(2-\pb) \;, \\
   I_7(\pb) &\equiv&\int x^3dx\, j_0(\pb x) \rhoSL(x)j_1(x) 
      = \frac{3\pi}{8\pb}(2-\pb^2)\,\theta(2-\pb) \;, \\
   I_8(\pb) &\equiv&\int x^3dx\, j_0(\pb x) \rhoSL(x)j_3(x) 
     \nonumber \\
     &=& \frac{3\pi}{32\pb}(-8 + 40\pb - 36\pb^2+5\pb^4)\,
       \theta(2-\pb) \;.
\eea

With explicit expressions for the DME approximation to $F_{1x}(\Rvec,\pvec,\qvec)$ in hand,
all that remains is to insert Eqs.~(\ref{eq:F1x_A})--(\ref{eq:F1x_C}) into
Eq.~(\ref{eq:W_c1x_jac}) and group terms accordingly. The $A[\rho]$ and
$B[\rho]$  coupling functions follow immediately and are given by
\bea
  A[\rho]^{1x}_{2\pi} 
    &=& -\frac{3g_A^2\rho^3}{16\pi f_\pi^2\kf ^3}
    \int q^2 dq\, V_{c_1c_3}(0,q) \, [I_1(\qb) + \frac{6}{5}I_2(\qb)] 
    \;, \\
  B[\rho]^{1x}_{2\pi} 
    &=& \frac{3g_A^2\rho^2}{8\pi f_\pi^2 \kf ^5}
     \int q^2dq\, V_{c_1c_3}(0,q) \, I_2(\qb)\;.
\eea


The derivation of the $C[\rho]_{2\pi}^{1x}$ coupling is a bit more complicated
because we must first partially integrate all $\nabla^2\rho$ terms.  Writing the
gradient contributions to $\WHF^{1x}$ as
\bea
    \WHF^{(1x)}\bigr|_{|\nabla\rho|^2} &=& 
   \int d\Rvec\,\bigl[ C_{\nabla^2\rho}^{1x}\nabla^2\rho(\Rvec)
   + C_{|\nabla\rho|^2}^{1x}|\nabla\rho(\Rvec)|^2\bigr] 
   \nonumber \\
   &=&  \int d\Rvec\,|\nabla\rho(\Rvec)|^2\,\bigl[ C_{|\nabla\rho|^2}^{1x} 
   -\frac{d}{d\rho}C_{\nabla^2\rho}^{1x}\bigr] \;,
\eea
we obtain
\beqn
    C[\rho]^{1x}_{2\pi} 
      = C_{|\nabla\rho|^2}^{1x} - \frac{d}{d\rho}C_{\nabla^2\rho}^{1x}\;.
\eeqn
Comparing to Eqs.~(\ref{eq:W_c1x_jac}) and (\ref{eq:F1x_C}) we find
\bea
   C_{|\nabla\rho|^2}^{1x} 
   &=& -\frac{g_A^2 \rho}{12\pi^2f_\pi^2 \kf ^8}\int p^2dp\,q^2dq\,
    \overline{V}_{c_1c_3}(p,q) \, I_3(\pb)
   \nonumber \\
   && \qquad\qquad\qquad \null \times
   \bigl\{ -I_6(\qb) -  \frac{2}{15}I_7(\qb) +\frac{1}{5}I_8(\qb)\bigr\}
\eea
and
\bea
   C_{\nabla^2\rho}^{1x} 
     &=& -\frac{3 g_A^2 \rho^2}{32\pi f_\pi^2 \kf ^5}\int q^2dq\,
    V_{c_1c_3}(0,q) \, I_2(\qb) 
  \nonumber \\
   &&
    \null -
   \frac{ g_A^2 \rho^2}{4\pi^2 f_\pi^2 \kf ^8}\int p^2dp\,q^2dq\,
   \overline{V}_{c_1c_3}(p,q)I_3(\pb)
   \bigl\{I_1(\qb) -\frac{1}{3}I_6(\qb)\bigr\}\;,
\eea
where the angle-averaged interaction $\overline{V}_{c_1c_3}(p,q)$ is defined as
\beqn
   \overline{V}_{c_1c_3}(p,q) \equiv 
   \frac{1}{2}\int d\cos\theta\, V_{c_1c_3}(\pvec,\qvec)\;.
\eeqn

\subsection{c-term double exchange}

The double exchange contribution from the c-term is given in
Eq.~(\ref{eq:HF2x}).  Since this involves a product of three
off-diagonal  density matrices, the DME is significantly more involved
than for the other 3N contributions. In order to assess the sensitivity
to the details  of the (non-unique) DME prescription, we consider two
different expansion schemes for these contributions, which we denote by
DME I  and DME II.  We expect the differences between the two schemes
should be  ``small'' if the master formula Eq.~\eqref{eq:DMEmastereqn}
is indeed a controlled expansion,  and if results are insensitive to the
different angle-averaging used in the two schemes.

\subsubsection{DME I}
We start by noting that repeated application of the master formula
Eq.~(\ref{eq:DMEmastereqn}) factorizes the three-body center-of-mass
and relative coordinate dependence as
\beq
\rho(\xvec_i,\xvec_j)=\sum_l \lambda_l(r_m,r_{ij})\mathcal{O}_l(R) \;.
\eeq
where $\mathcal{O}_{l}(R)$ is some monomial of the local densities and
$i,j,m$ are a permutation of 1, 2, and 3. The relative coordinate
functions can be written in terms of their Fourier transforms, e.g.,
\begin{equation}
\tilde{\lambda}(k_m,k_{ij})=\int \frac{d\rvec_m\,d\rvec_{ij}}{(2\pi)^6}
e^{i \kvec_m {\rvec_m}}e^{i \kvec_{ij} {\rvec_{ij}}}\lambda(r_m,r_{ij}) \;.
\end{equation}
Expanding the appropriate set of Jacobi coordinates for each density
matrix, Eq.~(\ref{eq:HF2x}) can therefore be written as
\begin{eqnarray}
\nonumber
\WHF^{(2x,c)}&=&\frac{1}{(2\pi)^{18}}\frac{3 g_c}{16}
\sum_{ijm}\int d\xvec_1\,d\xvec_2\,d\xvec_3
d\qvec_2\,d\qvec_3\,\mathcal{D}\kvec\:
V_{c_1c_3c_4}(\pvec,\qvec)
\\
\nonumber
&&\qquad \null\times 
\tilde{\lambda}_i(k_1,k_{23})
\tilde{\lambda}_j(k_2,k_{13})
\tilde{\lambda}_m(k_3,k_{12})
\mathcal{O}_i(R)\mathcal{O}_j(R) \mathcal{O}_k(R)\\
&& \null\times e^{-i(\mathbf{k_1 \cdot r_1} + \mathbf{k_2 \cdot r_2}
  +\mathbf{k_3 \cdot r_3} +\mathbf{ k_{13} \cdot r_{13}} 
  +\mathbf{k_{12} \cdot r_{12}} +\mathbf{k_{23} \cdot r_{23}}
  +\mathbf{q_{2} \cdot r_{12}}-\mathbf{ q_{3} \cdot r_{13}})} \;, 
\label{eq:W2x1}
\end{eqnarray}
with $g_c=(g_A/2 f_\pi)^2$ and where $\mathcal{D}\kvec$ denotes an
integration over all variables of type $\kvec_m$ and $\kvec_{ij}$ and
\begin{eqnarray}
V_{c_1c_3c_4}(\pvec,\qvec)&=& \frac{\qvec_2\cdot\qvec_3}
  {(q_2^2 + m_\pi^2)(q_3^2 + m_\pi^2)} \bigl[ - \frac{4 c_1 
  m_\pi^2}{f_\pi^2} + \frac{2 c_3}{f_\pi^2}\bigl(1 + c_4/c_3\bigr) 
    \: {\bf q}_2 \cdot {\bf q}_3 \bigr]
  \nonumber \\
  &&\hspace{45mm} - 2 \frac{c_4}{f_\pi^2}
       \frac{q_2^2q_3^2}{(q_2^2 + m_\pi^2)(q_3^2 + m_\pi^2)} \;.
\end{eqnarray}
Now choose one set of Jacobi coordinates, e.g., $\rvec_2$ and $\rvec_{13}$ and
rewrite Eq.~(\ref{eq:W2x1}) in terms of these alone
\bea
\nonumber
\kvec_1 \bfcdot \rvec_1&\longrightarrow& 
  \kvec_1 \bfcdot (-\frac{1}{2} \rvec_2-\frac{3}{4}
  \rvec_{13}) \;, 
\qquad\kvec_{23} \bfcdot \rvec_{23}
  \longrightarrow\kvec_{23} \bfcdot (\rvec_2-\frac{1}{2}\rvec_{13}) \;, \\
\nonumber
\kvec_3 \bfcdot \rvec_3
  &\longrightarrow& \kvec_3 \bfcdot (-\frac{1}{2} \rvec_2+\frac{3}{4}
\rvec_{13}) \;, 
\qquad\kvec_{12} \bfcdot\rvec_{12}
\longrightarrow\kvec_{12} \bfcdot(-\rvec_2-\frac{1}{2}\rvec_{13}) \;, \\
\qvec_2 \bfcdot \rvec_{12}&
  \longrightarrow& \qvec_2 \bfcdot(-\rvec_2-\frac{1}{2}\rvec_{13}) \;.
\eea
We obtain as our final result
\bea
\label{eq:W2x-final}
\nonumber
\WHF^{(2x,c)}&=&\frac{3 g_c}{16 (2\pi)^{18}}\sum_{i,j,m}\int d \Rvec \,\mathcal{D} \kvec\,
\tilde{\lambda}_i(k_1,k_{23})
\tilde{\lambda}_j(k_2,k_{13})
\tilde{\lambda}_m(k_3,k_{12})
\\
&&\hspace{30mm}
\times\mathcal{O}_i(R)\,\mathcal{O}_j(R)\, \mathcal{O}_k(R)
\,V_{c_1 c_3 c_4}(\mathcal{K}_1,\mathcal{K}_2) \;, 
\eea
with
\begin{eqnarray}
\nonumber
  \label{eq:2xmomenta}
  \mathcal{K}_1&=&\kvec_2-\frac{1}{2}\kvec_1-\frac{1}{2}\kvec_3+\kvec_{23}-\kvec_{12} \;, \\
  \mathcal{K}_2&=&\kvec_{13}-\frac{1}{2}\kvec_1+\kvec_3-\kvec_{23} \;.
\end{eqnarray}
Now let us consider the particular form of the functions appearing
in the integrals. We expand each density matrix as in
Eq.~(\ref{eq:dme_single_matrix}) and use Eq.~(\ref{eq:del-alpha})
to evaluate the $\nabla^2(\rho_{SL}\,\rho)$ term:
\bea
\nonumber
\rho(\xvec_1,\xvec_2)&\approx&\rho\bigl[\rho_{\rm SL}(k_F r_{12})+r_{12}^2 g(k_F
r_{12})\frac{3}{5}k_F^2\bigr]+\tau\bigr[-r_{12}^2 g(k_F r_{12})\bigr]\\
\nonumber&&
\qquad+\nabla^2\rho\bigl[\frac{r_3^2}{9}g(k_F r_3)j_0(k_F r_{12})
+\frac{r_{12}^2}{4}g(k_F r_{12})\bigr]\\
&&\qquad+|\nabla \rho|^2\bigr[-\frac{r_3^2}{9}g(k_F r_3)
\frac{(k_F r_{12})^2}{9 \rho}\rho_{\rm SL}(k_F r_{12})\bigr] \;.
\eea
This leads us to define
\bea
\lambda_1(r_3,r_{12})&\equiv&\left(\rho_{\rm SL}(k_F r_{12})+r_{12}^2 g(k_F r_{12})\frac{3}{5} k_F^2\right) \;, \\
\lambda_2(r_3,r_{12})&\equiv&-r_{12}^2 g(k_F r_{12}) \;, \\
\lambda_3(r_3,r_{12})&\equiv&\frac{r_3^2}{9}g(k_F r_3)j_0(k_F r_{12})
+\frac{r_{12}^2}{4}g(k_F r_{12}) \;, \\
\lambda_4(r_3,r_{12})&\equiv&-\frac{r_3^2}{9}g(k_F r_3)
\frac{(k_F r_{12})^2}{9 \rho}\rho_{\rm SL}(k_F r_{12}) \;.
\eea
We obtain the A-term by inserting the relevant functions into
Eq.~(\ref{eq:W2x-final})
\beq
A[\rho] = \frac{3g_c\,\rho^3}{16(2\pi)^{18}}\int\mathcal{D}\kvec\,
\tilde{\lambda}_1(k_3,k_{12})\tilde{\lambda}_1(k_2,k_{13})
\tilde{\lambda}_1(k_1,k_{23}) V_{c_1 c_3 c_4}(\mathcal{K}_1,\mathcal{K}_2)
 \;,
\eeq
with
\bea
\nonumber
\tilde{\lambda}_1(k_3,k_{12})&=&\left(\frac{6\pi^2}{k_F^3}+\frac{21\pi^2}{2
    k_F^5}(3k_F^2-5k_{12}^2) \right)
\Theta(k_F-k_{12})(2\pi)^3\delta^{(3)}(\kvec_{3})\\
&=&\tilde{\lambda}_1(k_{12})(2\pi)^3\delta^{(3)}(\kvec_{3}) \;.
\eea
Integrating over the $\delta$-functions leads to
\bea
A[\rho] &=& \frac{3 g_c \,\rho^3}{16 (2\pi)^9}\int\,d\kvec_{12}\,d\kvec_{13}\,d \kvec_{23}\,
\tilde{\lambda}_1(k_{12})\tilde{\lambda}_1(k_{13})
\tilde{\lambda}_1(k_{23}) \nonumber \\
  & & \qquad\qquad \null \times V_{c_1 c_3 c_4}(\kvec_{23}-\kvec_{12},
\kvec_{13}-\kvec_{23})
 \;.
\eea
The B-term is proportional to $\tau$
\bea
\nonumber
B[\rho]&=&\frac{3 g_c\,\rho^2}{16(2\pi)^{18}}\int\,\mathcal{D}\kvec
V_{c_1 c_3 c_4}(\mathcal{K}_1,\mathcal{K}_2)
\biggl(
\tilde{\lambda}_2(k_3,k_{12})\tilde{\lambda}_1(k_2,k_{13})\tilde{\lambda}_1(k_1,k_{23})\\
&&
\nonumber
+\tilde{\lambda}_1(k_3,k_{12})\tilde{\lambda}_2(k_2,k_{13})\tilde{\lambda}_1(k_1,k_{23})
+\tilde{\lambda}_1(k_3,k_{12})\tilde{\lambda}_1(k_2,k_{13})\tilde{\lambda}_2(k_1,k_{23})\biggr) \;, \\
\eea
with
\bea
\nonumber
\tilde{\lambda}_2(k_3,k_{12})&=&-\frac{35\pi^2}{2
  k_F^7}(3k_F^2-5k_{12}^2)\Theta(k_F-k_{12})(2\pi)^3\delta^{(3)}(\kvec_3)\\
&=&\tilde{\lambda}_2(k_{12})(2\pi)^3\delta^{(3)}(\kvec_3) \;.
\eea
Integrating out the $\delta$-functions gives
\bea
\nonumber
B[\rho]&=&\frac{3 g_c\,\rho^2}{16(2\pi)^9}\int\,d\kvec_{12}\,d\kvec_{13}\,d\kvec_{23}
V_{c_1 c_3 c_4}(\kvec_{23}-\kvec_{12},\kvec_{13}-\kvec_{23})\\
\nonumber
&&\hspace{15mm}\times
\biggl(\tilde{\lambda}_2(k_{12})\tilde{\lambda}_1(k_{13})
\tilde{\lambda}_1(k_{23})+\tilde{\lambda}_1(k_{12})\tilde{\lambda}_2(k_{13})
\tilde{\lambda}_1(k_{23})\\
&&\hspace{60mm}
+\tilde{\lambda}_1(k_{12})\tilde{\lambda}_1(k_{13})\tilde{\lambda}_2(k_{23})\biggr) \;, 
\eea

The calculation of the relevant integrals for the $C$-term is more involved.
We first consider on the integral for the coefficient of $|\nabla \rho|^2$
\bea
\label{eq:C1}
\nonumber
C_{|\nabla \rho|^2}&=&\frac{3 g_c\,\rho^2}{16(2\pi)^{18}}\int\mathcal{D}\kvec\,
V_{c_1 c_3 c_4}(\mathcal{K}_1,\mathcal{K}_2)
\bigl(
\tilde{\lambda}_1(k_3,k_{12})\tilde{\lambda}_1(k_2,k_{13})\tilde{\lambda}_4(k_1,k_{23})\\
\nonumber
&&
+\tilde{\lambda}_1(k_1,k_{23})\tilde{\lambda}_1(k_2,k_{13})\tilde{\lambda}_4(k_3,k_{12})
+\tilde{\lambda}_1(k_1,k_{23})\tilde{\lambda}_1(k_3,k_{12})\tilde{\lambda}_4(k_2,k_{13})
\bigr) \;, \\
\eea
with
\bea
\nonumber
\tilde{\lambda}_4(k_1,k_{23})&=&-\frac{35 \pi^4}{81 \rho k_F^7}
(3k_F^2-5 k_1^2)\Theta(k_F-k_1)\\
&&\hspace{10mm}\times\left(\frac{1}{k_F k_{23}}\delta(k_F-k_{23})
+\frac{1}{k_{23}}\frac{\hbox{d}}{\hbox{d}k_{23}}\delta(k_F-k_{23})\right) \;.
\eea
Let us focus our attention on the first term in Eq.~(\ref{eq:C1}). This
term contains a factor
\beq
\biggl(\frac{1}{k_F}\delta(k_F-k_{23})
+\frac{\hbox{d}}{\hbox{d} k_{23}}\delta(k_F-k_{23})\biggr)
\frac{1}{k_{23}}V_{c_1 c_3 c_4} \;, 
\eeq
which simplifies to
\beq
\frac{2}{k_{23}k_F}\delta(k_F-k_{23})V_{2X}-\frac{1}{k_{23}}\delta(k_F-k_{23})\frac{\hbox{d}}{\hbox{d}k_{23}}V_{2X} \;.
\eeq
after partial integration.

The second term has the coefficient $\nabla^2\rho$
\bea
\nonumber
C_{\nabla^2\rho}&=&\frac{3g_c\,\rho^2}{16(2\pi)^{18}}\int\mathcal{D}^3k\,
V_{c_1 c_3 c_4}(\mathcal{K}_1,\mathcal{K}_2)
\bigl(
\tilde{\lambda}_1(k_3,k_{12})\tilde{\lambda}_1(k_2,k_{13})\tilde{\lambda}_3(k_1,k_{23})\\
\nonumber
&&+
\tilde{\lambda}_1(k_1,k_{23})\tilde{\lambda}_1(k_2,k_{13})\tilde{\lambda}_3(k_3,k_{12})+
\tilde{\lambda}_1(k_1,k_{23})\tilde{\lambda}_1(k_3,k_{12})\tilde{\lambda}_3(k_2,k_{13})
\bigr) \;,\\
\eea
with
\bea
\nonumber
\tilde{\lambda}_3(k_1,k_{23})&=&\frac{35\pi^4}{9 k_F^8 k_{23}}
(3 k_F^2-5k_1^2)\delta(k_F-k_{23})\Theta(k_F-k_1)\\
\nonumber
&&\qquad\qquad+ \frac{35\pi^5}{k_F^7}\delta^{(3)}(\kvec_1)
(3k_F^2-5k_{23}^2)\Theta(k_F-k_{23})\\
&=&\lambda_{3A}(k_1)\frac{1}{k_{23}}\delta(k_F-k_{23})
+\lambda_{3B}(k_{23})(2\pi)^3\delta(\kvec_1) \;.
\eea
Integrating over the $\delta$-functions leads to a lengthy expression
that we will not give here.

Using partial integration we can finally write the total expression
in the form
\beq
C[\rho]=C_{|\nabla \rho|^2}-\frac{\hbox{d}}{\hbox{d}\rho}C_{\nabla^2\rho}
  \;.
\eeq
The particular order of integrations we have carried out gives factors of $k_F$ 
appearing as UV cutoffs in the remaining integrals. Such a simplification 
arises for all contributions to the HF energy and
the resulting integrals can therefore be easily integrated numerically
despite the relatively large number of integration variables. 

Key to the prescription used here is the Fourier transform of the
expanded density matrices to momentum space. Due to its generality,
this approach
can easily be extended to the calculation of higher-order
contributions to the DME. A similar approach was introduced in 
Ref.~\cite{Kaiser:2002jz},
where the authors used the Fourier transform of the expanded density
matrix to generate medium insertions for a diagrammatic calculation of
the nuclear energy density functional using chiral perturbation
theory.

\subsubsection{DME II}
The DME I prescription outlined above differs from the original  NV
approach in two respects. First, we do not rearrange and truncate the 
expansion by hand to ensure that the nuclear matter limit is exactly
reproduced.  Second, the DME I prescription keeps cross-terms in the
product of the three expanded density  matrices that are formally of
higher order in the NV approach.  In order to quantify these effects and
assess whether the expansion is under control, we have also performed the 
expansion where we strictly follow the original NV philosophy (DME II).

We also note the differences in angle-averaging that
arise with the different DME schemes.  In the DME I approach, each
$\rho(\xvec_i,\xvec_j)$  is first expanded in the natural Jacobi
coordinates $(\Rvec, \rvec_k, \rvec_{ij})$, and then the three  expanded
density matrices are expressed in one common set of Jacobi coordinates.
In the DME II prescription, we follow a different path by expressing the
product of density matrices in  one common set of Jacobi coordinates
from the outset. The subsequent DME implies a different 
angle-averaging, since only one density matrix is expanded in its
natural Jacobi basis.  We do not include the derivation of the DME II
equations here, as it proceeds in much the same spirit as for the DME I,
although we note that the final expressions are considerably more
cumbersome since one finds different $\widetilde{\lambda}_l$ functions
depending on whether one is expanding the $\rho(\xvec_{i},\xvec_{j})$
corresponding to the chosen Jacobi coordinates or one of the other two
density matrices.  
\section{Results}
 \label{sec:results}

In this section, we make some basic tests of the DME.
We have two modest goals: to check that the DME does not degrade when
applied to non-local, low-momentum NN potentials and to make
a first assessment of the relative contributions of two- and
three-body interactions.
For the first goal, we approximate the
self-consistent Hartree-Fock ground-state wave function by a Slater
determinant of harmonic oscillator single-particle wave functions. 
Using these wave functions, we compare the DME approximation for
the energy of a schematic model NN potential
to the exact result where the finite range and
non-locality of the interaction is treated without approximation.
Then with the same wave function we check the error as we
change the resolution (cutoff) of a realistic low-momentum potential.
For the second goal, we exhibit some numerical results for the DME
coefficient functions
to illustrate the non-trivial density dependence and to show the
effects of different prescriptions for the three-body DME. 
These are meant only to set a
baseline because, at a minimum, we should include second-order
contributions (i.e., beyond Hartree-Fock) before expecting
quantitative predictions for nuclear structure or analyzing the cutoff
dependence of the energy functional. However, even at this stage it
should be meaningful to use these results to compare  the relative
contributions of two- and three-body interactions.

\begin{figure}[pt]
 \begin{center}
  \includegraphics*[width=3.5in]{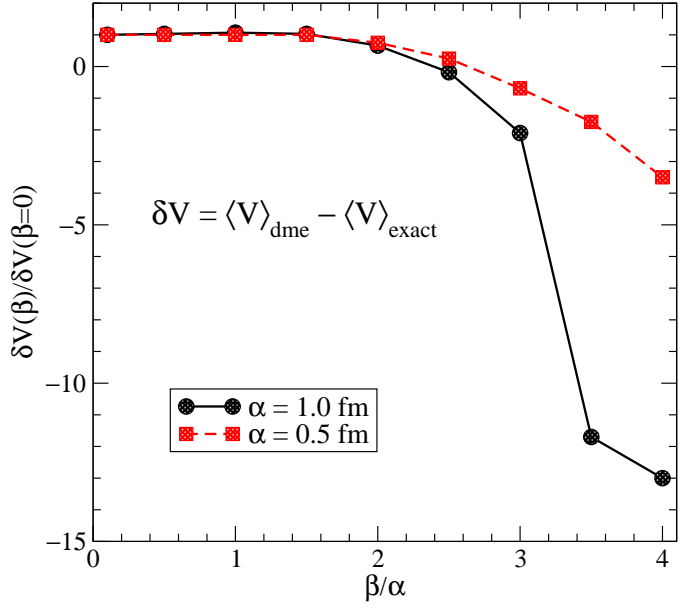}
  
  
 \end{center}
 \caption{Effects of different non-localities on the accuracy
 of the DME as a function of the ratio of the non-locality to
 range parameters for the harmonic oscillator approximation
 to the ground state of $^{40}$Ca.}
 \label{fig:deltaV}
\end{figure}

Although the original DME paper introduced formalism for non-local
potentials~\cite{NEGELE72},
previous investigations of the effectiveness of the
DME studied only local potentials (or local approximations to
the G~matrix).
Because the low-momentum potentials used here can be strongly
non-local, we first test whether the extra expansion required
degrades the accuracy of the DME.
We consider a model potential:
\beqn
  V(\rvec,\rvec') = v\Bigl(\frac{\rvec+\rvec'}{2\alpha}\Bigr)
      \,
      \frac{1}{(\pi\beta^2)^{3/2}}\, e^{-(\rvec-\rvec')^2/\beta^2}
      \;,
    \label{eq:modelV}  
\eeqn 
with $v$ a Gaussian potential, so the range is set by $\alpha$.  
The range of the non-locality is set by $\beta$; in the limit
$\beta \rightarrow 0$, $V(\rvec,\rvec') \rightarrow
v(\rvec/\alpha)\delta^3(\rvec-\rvec')$.

 In Fig.~\ref{fig:deltaV}, the effects of non-localities 
on the accuracy of the DME for integrated quantities 
(e.g., $\langle V\rangle$) is illustrated using this potential.
We use a harmonic oscillator model of $^{40}$Ca (i.e.,
the ground-state wave function is a Slater determinant of harmonic oscillator
orbitals) and calculate the expectation value of the non-local 
$V(\rvec,\rvec')$ in the Hartree-Fock
ground state.
For a given range $\alpha$, we compare the error for a non-locality
$\beta$ to the error with $\beta = 0$.  It is evident that the effect
of the non-locality on the degradation of the DME is unimportant up to at least twice the range.
Even when $\alpha$ is taken as small as the typical range of a 
repulsive core there should be no problem for the range of low-momentum 
cutoffs typically considered. 

\begin{figure}[t]
  \begin{center}
  \includegraphics*[width=3.8in]{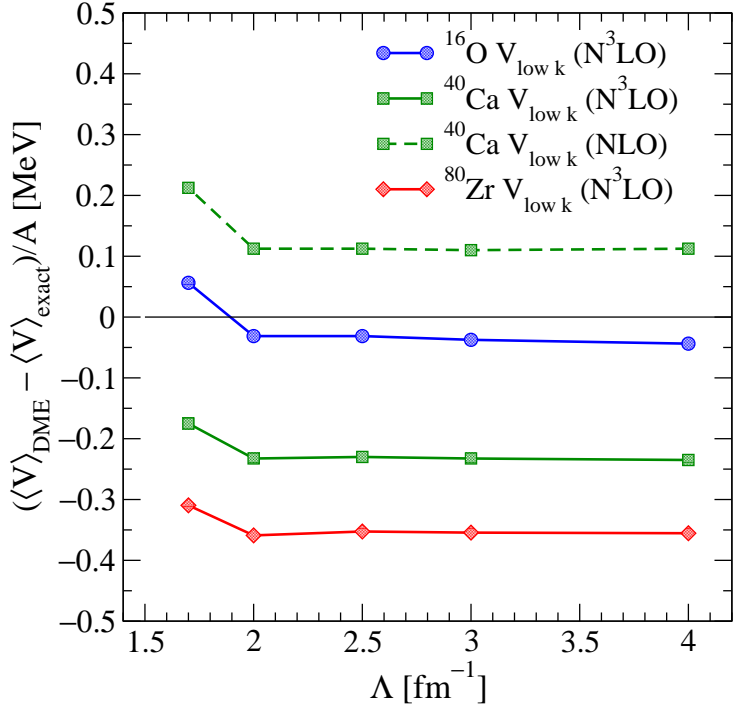}
  \end{center}
  \caption{Errors per nucleon in the DME predictions for the expectation
  value of a model potential, Eq.~(\ref{eq:modelV}), 
  in a harmonic oscillator ground state for
  three $N=Z$ nuclei (no Coulomb interaction).}
  \label{fig:DMEerrorA}
\end{figure}

The errors per nucleon for the DME with the same model ground state 
but with a realistic low-momentum nucleon-nucleon 
potential (starting from the chiral N$^3$LO potential
from Ref.~\cite{N3LO}) are shown in Fig.~\ref{fig:DMEerrorA} for
$N=Z$ nuclei (without Coulomb) for $A = 16$, 40, and 80.
It is evident that the cutoff dependence of the
error is very slight until $\Lambda < 2\,\mbox{fm}^{-1}$.
Because the evolution of the potential does not alter the long-distance
part, the weak cutoff dependence of the error implies 
that the short-distance contribution is very
well reproduced and provides further confirmation that non-locality (which
grows with decreasing $\Lambda$) is not a problem for the DME
(note that long-range local interactions remain local).
These errors are also smaller than errors found in early DME
tests.

The model calculations in Fig.~\ref{fig:DMEerrorA} treat both direct (Hartree)
and exchange terms with the DME.  It was recognized long ago that
the DME is ill-suited for long-range direct terms, which should be 
calculated exactly instead~\cite{NEGELE75}.  
The dashed line in the figure shows the
error for $A=40$ but using the NLO potential, which does not have
any long-range contributions to the direct scalar term.  
As expected, the
error is significantly smaller than the N$^3$LO result, due at least
in part to the crude treatment of the N$^3$LO long-range direct
contribution.
Since the long-range local terms can be isolated in the potential, 
it is feasible
to perform exact Hartree evaluations of these pieces when implemented
in a DFT solver.

\begin{figure}[tp]
 \begin{center}
  \includegraphics*[width=4.0in]{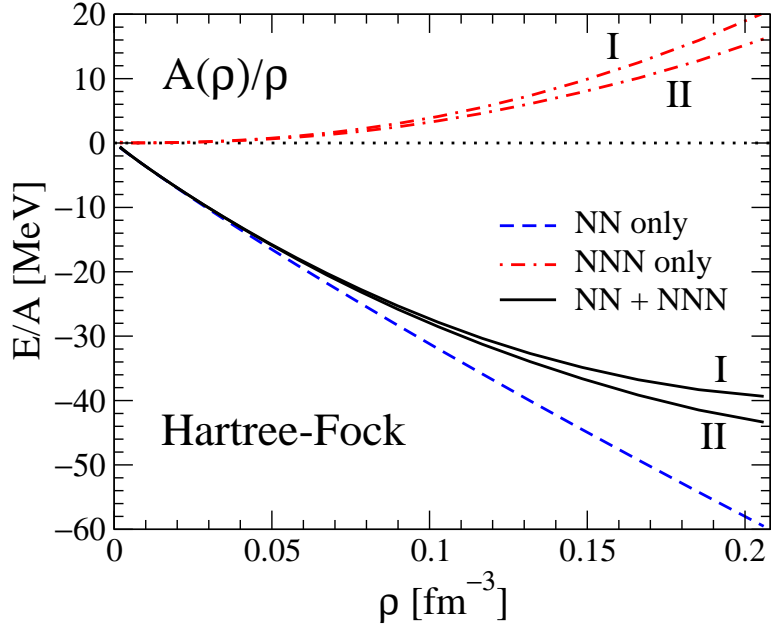}
 \end{center}
 \vspace*{-.1in}
 \caption{Contribution to the energy per particle in nuclear
 matter from the isoscalar coefficient function $A(\rho)$ 
 as a function of the density from the DME applied to the Hartree-Fock
 energy calculated using $\vlowk$ with $\Lambda = 2.1\fmi$.
 The result including the NN interaction alone is compared
 to NN plus NNN interactions for two DME expansions (I and II, see text).}
 \label{fig:A_coeff}
\end{figure}

\begin{figure}[tp]
 \begin{center}
  \includegraphics*[width=4.0in]{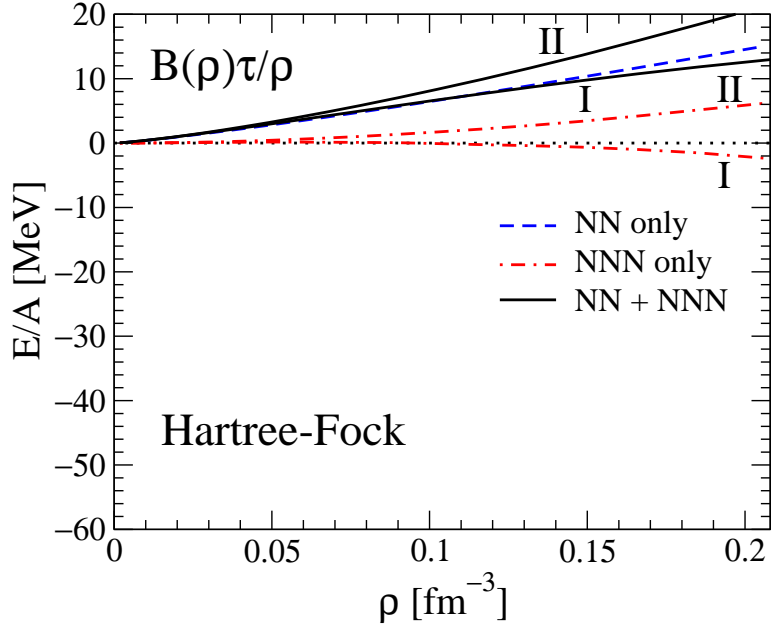}
 \end{center}
 \vspace*{-.1in}
 \caption{Contribution to the energy per particle in nuclear
 matter from the isoscalar coefficient function $B(\rho)$ 
 as a function of the density from the DME applied to the Hartree-Fock
 energy calculated using $\vlowk$ with $\Lambda = 2.1\fmi$.
 The result including the NN interaction alone is compared
 to NN plus NNN interactions for two DME expansions (I and II, see text).}
 \label{fig:B_coeff}
\end{figure}

\begin{figure}[tp]
 \begin{center}
  \includegraphics*[width=4.0in]{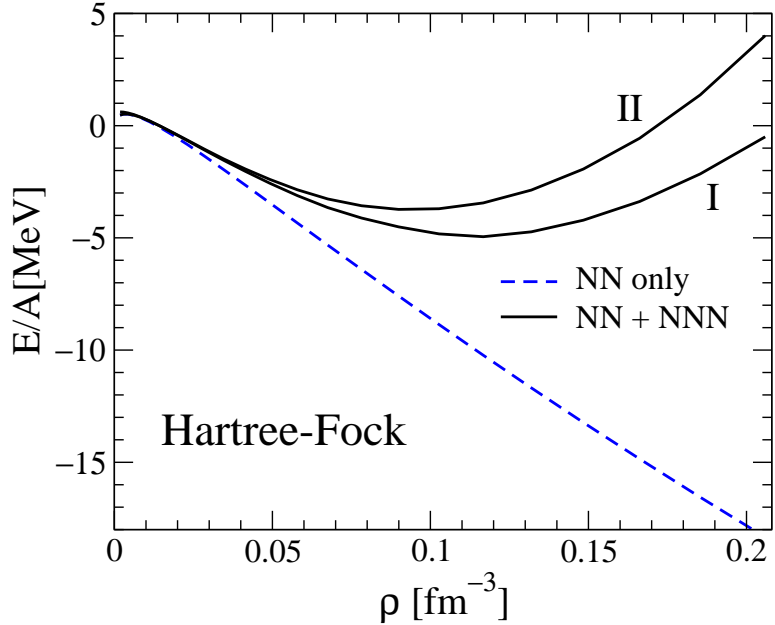}
 \end{center}
 \vspace*{-.1in}
 \caption{Energy per particle in nuclear
 matter by combining contributions from
 from the isoscalar coefficient functions $A(\rho)$ and $B(\rho)$
 with the kinetic energy 
 as a function of the density from the DME applied to the Hartree-Fock
 energy calculated using $\vlowk$ with $\Lambda = 2.1\fmi$.
 The result including the NN interaction alone is compared
 to NN plus NNN interactions for two DME expansions (I and II, see text).
 Note that only expansion II correctly reproduces the nuclear matter limit.}
 \label{fig:AB_coeff}
\end{figure}

We turn now to the isoscalar $A$ and $B$ functions, which are the only
contributors to uniform, symmetric nuclear matter.
The energy per particle as a function of density $\rho$ is given by:
\beqn
  E/A = \frac{1}{\rho}\left[
         \frac{\hbar^2}{2M}\tau
        + A(\rho) + B(\rho)\tau
     \right]
     \ .
     \label{eq:EoverA}
\eeqn
The individual contributions from $A$ and $B$ at the Hartree-Fock level
are plotted in Figs.~\ref{fig:A_coeff}
and \ref{fig:B_coeff}, and combined into $E/A$ in Fig.~\ref{fig:AB_coeff}.
These use a two-body $\vlowk$ interaction evolved from 
the Argonne $v_{18}$ potential~\cite{AV18} with
a sharp cutoff at $\Lambda = 2.1\,\fmi$ and a chiral N$^2$LO
three-body force with constants fit to the binding energies of
 the triton and $^4$He~\cite{Vlowk3NF}.
Results are given using the NN contribution only and with NNN included,
using the two prescriptions for the NNN double-exchange contribution
(DME-I and DME-II) described in Section~\ref{sec:dme-three-body}.
 
From Figs.~\ref{fig:A_coeff} and \ref{fig:B_coeff}, one sees that
the ratios of contributions from three-body to two-body tend to increase
monotonically with density, but are still only about 20--30\% at saturation
density.
This is consistent with general
expectations from chiral power counting. 
The actual scaling with density of the ratio varies only 
slightly from being
linear in the density.
Because the local density in actual nuclei in somewhat lower, there
is reason to believe the expansion in many-body forces is 
under control.  
Past estimates of contributions to Skyrme energy functionals based on naive
dimensional analysis~\cite{Furnstahl:1997hq} suggested large contributions
from three-body and even four-body interactions.
The present results imply more modest contributions, 
but evaluating the chiral N3LO
four-body contribution at Hartree-Fock will be needed for a definitive
assessment.

The comparison of the DME-I and DME-II curves gives us an estimate of
the truncation error in the expansion applied to the NNN
terms because these prescriptions
differ in the contributions of higher-order terms in the expansion.
Indeed, we have verified that suppressing these terms by hand brings
the predictions for the $A$ and $B$ coefficients into agreement.
The qualitative difference for the NNN-only contribution to $B$ is
large, but the actual coefficient itself is small, so this should
not be alarming.  However, because the combination of $A$ and $B$ and 
the kinetic energy to obtain the nuclear matter energy per particle
involves strong cancellations, the spread in Fig.~\ref{fig:AB_coeff}
is large on the scale of nuclear binding energies.

These differences motivate a generalization of the Negele-Vautherin
DME following the discussion in Ref.~\cite{Dobaczewski:2003cy}.
In this approach, the expansion of the scalar density matrix
takes the factorized form
\beq
  \rho(\Rvec + \frac{\svec}{2}, \Rvec - \frac{\svec}{2})
    = \sum_n \Pi_n(\kf s) \langle \mathcal{O}_n(\Rvec) \rangle
    \;,
\eeq
where 
\beq
   \langle \mathcal{O}_n(\Rvec) \rangle = 
   \{
    \rho(\Rvec), \tau(\Rvec), \nabla^2\rho(\Rvec), \cdots
   \}
   \;,
\eeq
and $\kf$ is a momentum scale typically taken to be
$\kf(\Rvec)$ as in Eq.~\eqref{eq:standardLDA}.
Similar expansions are made for the other components of the density
matrix.
Input from finite nuclei can be used to determine the $\Pi_n$ functions,
which can be viewed as general resummations of the DME expansion;
see Section~\ref{sec:summary} for a brief overview.  
 
\begin{figure}[tp]
 \begin{center}
  \includegraphics*[width=4.0in]{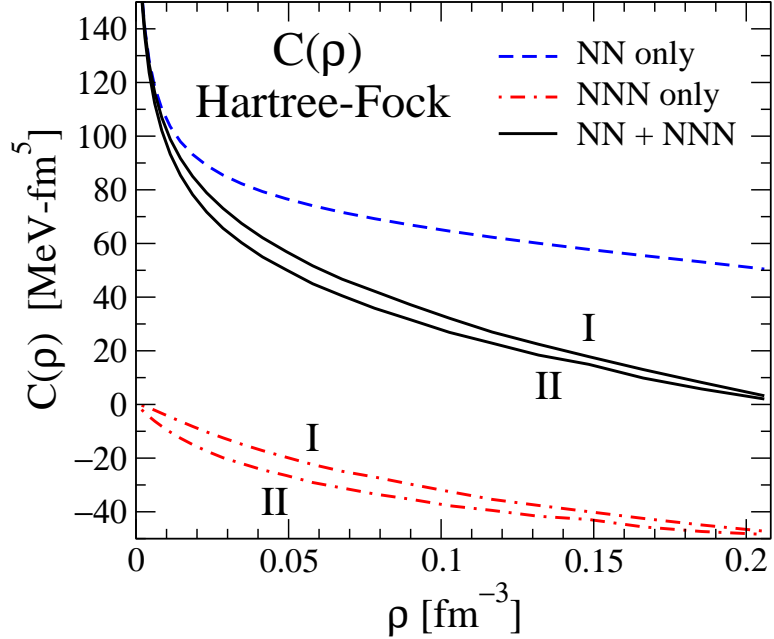}
 \end{center}
 \vspace*{-.1in}
 \caption{Isoscalar coefficient function $C(\rho)$ as a function of the 
 density from the DME applied to the Hartree-Fock
 energy calculated using $\vlowk$ with $\Lambda = 2.1\fmi$.
 The result including the NN interaction alone is compared
 to NN plus NNN interactions for two DME expansions (I and II, see text).}
 \label{fig:C_coeff}
\end{figure}

Finally, in Fig.~\ref{fig:C_coeff}, the coefficient function $C(\rho)$
is plotted as a function of density ($\rho = 2\kf^3/3\pi^2$).  Even at
the highest density, the three-body contribution is a manageable
correction to the two-body result. The NN + NNN result  is in
qualitative  agreement with the results of Fritsch and collaborators
who included two-pion exchanges with explicit
$\Delta$-isobars~\cite{Kaiser:2005},  although the three-body
contributions in the current work are somewhat larger than effects
arising from explicit $\Delta$-isobars.
For this coefficient function the difference between DME-I and DME-II
is comparatively small.


\section{Summary}
  \label{sec:summary}
  
\begin{figure}[b]
 \begin{center}
  \includegraphics*[width=2.5in]{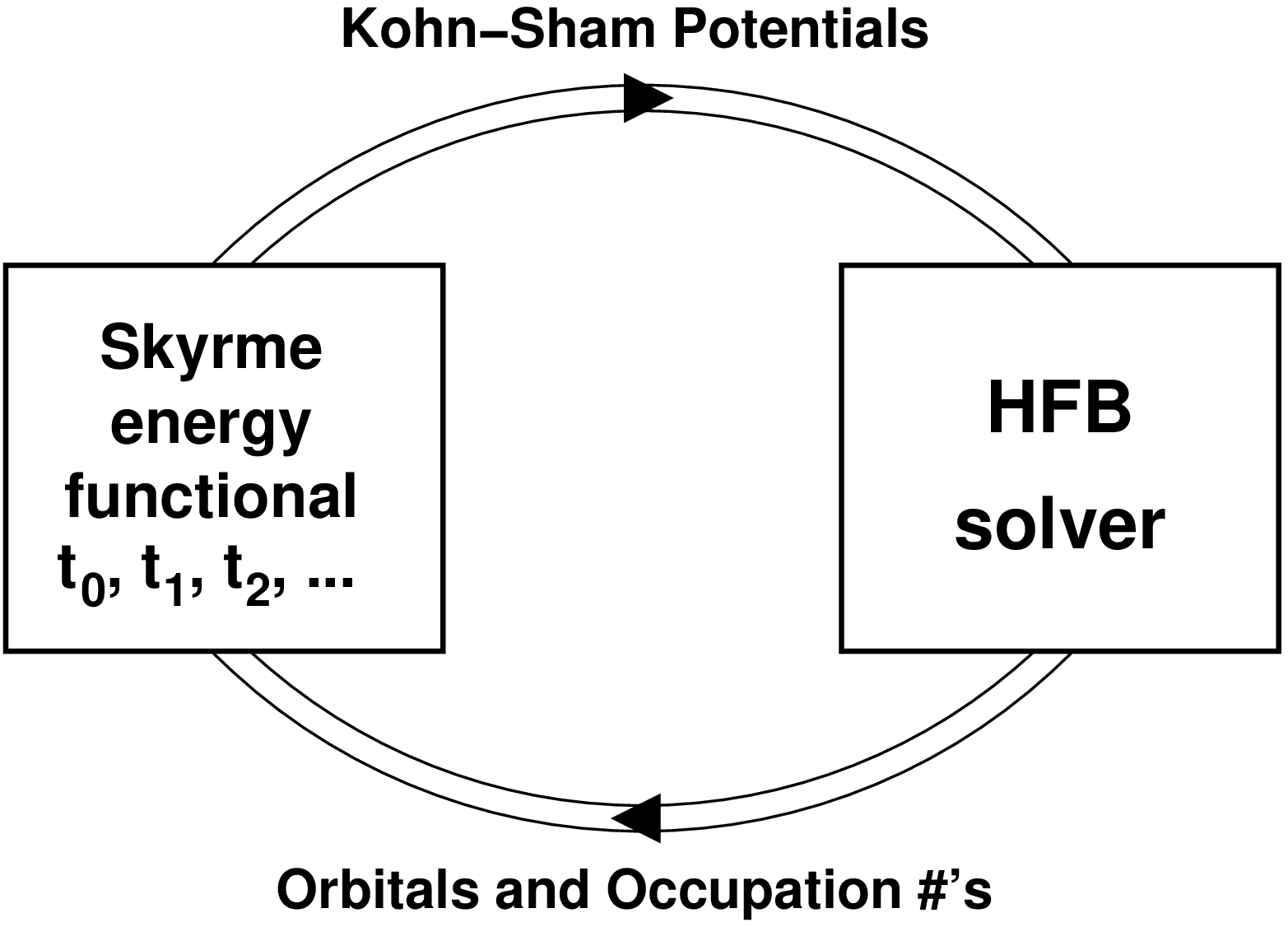}
  \hspace*{.1in}
  \includegraphics*[width=2.5in]{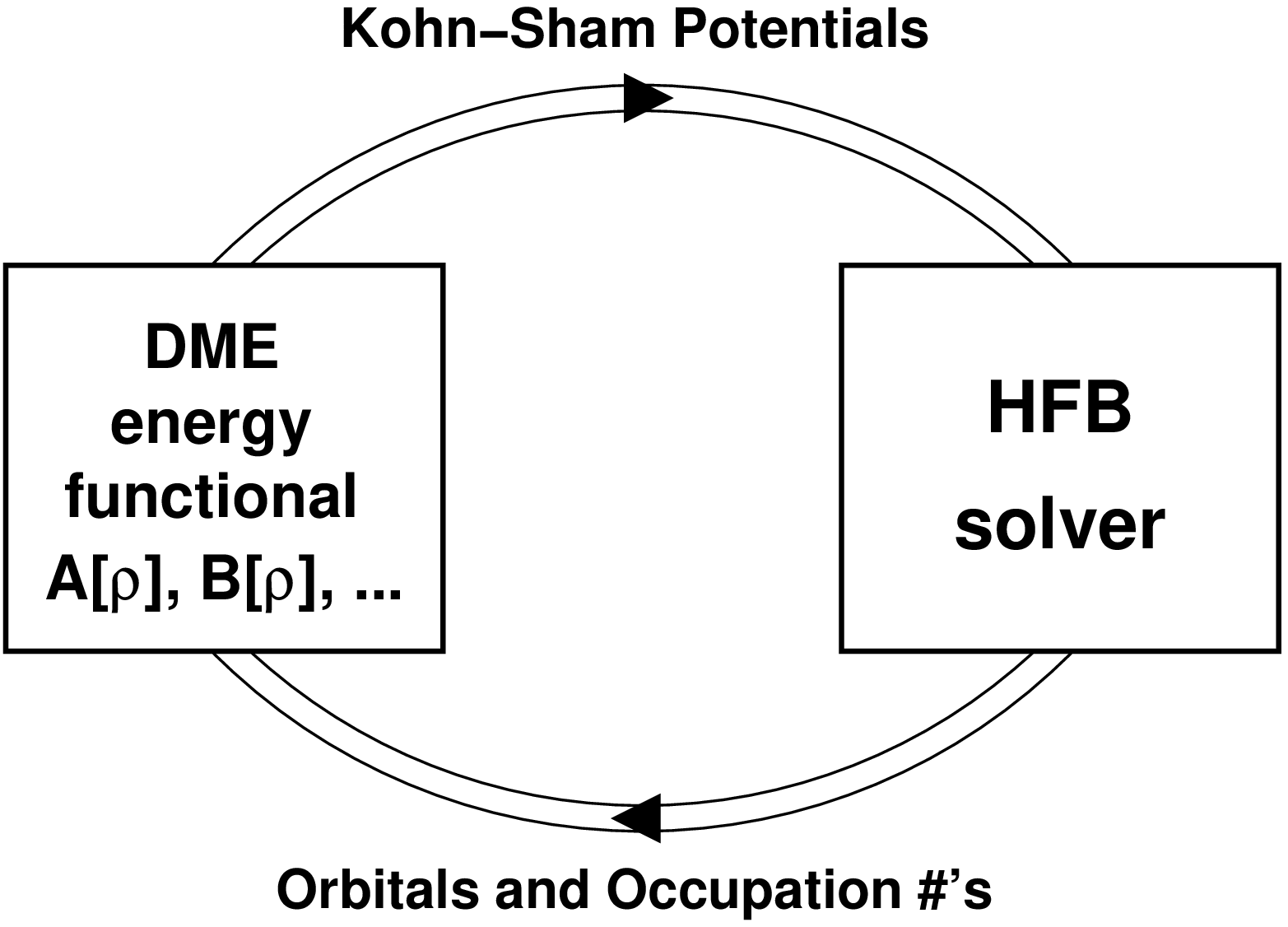}
 \end{center}
 \caption{Diagrams showing the flow in Skyrme HF codes at present
 (left) and modified for the DME (right).}
 \label{fig:dft_diagrams}
\end{figure}

In this paper, we have formulated the density matrix expansion 
(DME) for low-momentum
interactions and applied it to a Hartree-Fock energy functional including
both NN and NNN potentials.
The output is a set of functions of density that can replace 
density-independent
parameters in standard Skyrme Hartree-Fock energy density functionals.
This replacement in Skyrme HF computer codes
is shown schematically in Fig.~\ref{fig:dft_diagrams}. 
Only one section of such a code would be replaced, and it takes the same inputs
(single-particle eigenvalues
and wave functions for the orbitals and the corresponding
occupation numbers) and delivers the same outputs (local Kohn-Sham
potentials).  
Furthermore, the upgrade from Skyrme energy functional to DME energy
functional can be carried out in stages.  For example, the spin-orbit
part and pairing can be kept in Skyrme form with the rest given by 
the DME. 
Details of such a DME implementation will be given elsewhere.
A further upgrade to orbital-based methods would also only modify the same
part of the code, although the increased computational load will
be significant.
 
The numerical
results given here are limited and do not touch on many of the most
interesting aspects of microscopic DFT from low-momentum potentials.
Topics to explore in the future include:
\begin{itemize}
  \item Examine the resolution or scale dependence of the energy functional
  by evolving the input low-momentum potential.  There will be
  dependence on the cutoff $\Lambda$ (if using $\vlowk$) or the flow
  parameter $\lambda$ (if using $\vsrg$) both from omitted physics
  and from intrinsic scale dependence.  Calculations 
  at least to second order are needed to separate these dependencies.
  \item Examine the isovector part of the functional. We can isolate the contributions
  from the more interesting long-range (pion) parts of the free-space interactions, allowing
  us to obtain analytic expressions for the dominant density dependence of the isovector DME coupling functions.   
    \item Study the dependence of spin-orbit contributions on NN
  vs.\ NNN interactions.  This includes the isospin dependence as
  well as overall magnitudes. The NN spin-orbit contributions arise from 
  short-range interactions, whereas NNN contributions arise from 
  the long-range two-pion exchange interaction.Therefore, we expect to find a rather
  different density dependence for the two types of spin-orbit contributions.   
  \item Explore the contribution of tensor contributions, which
  have recently been reconsidered phenomenologically~\cite{Lesinski:2007,Brink:2007}.
  \item Understand the scaling of contributions from many-body forces.
  In particular, how does the four-body force (which is
  known at N$^3$LO
  in chiral EFT with conventional Weinberg counting) contribution
  at Hartree-Fock level impact the energy functional?  
  \end{itemize}

The calculations presented here are only the first step on the road to
a universal nuclear energy density functional (UNEDF)~\cite{UNEDF}. 
There are both refinements within the DME framework and generalizations that
test its applicability and accuracy. 
While many of these steps offer significant challenges, in every
case a plan is in hand to carry it out.
The DME can be directly extended to include second-order (or full
particle-particle ladder) contributions
by using averaged energies for the energy denominators.
However, a more systematic approximation is under development using
a short-time expansion~\cite{Vincent:2007}. 
More difficult future steps include dealing with
symmetry breaking and restoration in DFT for self-bound systems,
dealing with non-localities from near-on-shell particle-hole
excitations (vibrations), and incorporating pairing in the
same microscopic framework (see Ref.~\cite{Duguet:2007be}).

In extending our calculations we will also modify
the standard DME formalism from Ref.~\cite{NEGELE72} that we have followed
in the present work.
The formalism has problems even beyond the truncation errors
from different DME prescriptions
already discussed in Sections~\ref{sec:dme-three-body}
and \ref{sec:results}, 
the most severe being that
it provides an extremely poor description of the vector part of the density
matrix. 
While the standard DME is better at reproducing the scalar
density matrices, even here the errors are sufficiently large that the
disagreement with a full finite-range Hartree-Fock calculations can 
reach the MeV per particle level. 
Gebremariam and collaborators have
traced both of these problems to an
inadequate phase space averaging (PSA) used in the previous DME
approaches \cite{Biruk}. In the derivation of the DME, one incorporates average
information about the local momentum distribution into the approximation.
The Negele-Vautherin DME uses the phase space of infinite nuclear
matter to perform this averaging. However, the local momentum distribution
in finite Fermi systems exhibits two striking differences from that of
infinite homogenous matter. First, mean-field calculations of nuclei show
that the local momentum distribution exhibits a diffuse Fermi surface that
is especially pronounced in the nuclear surface. Second, the local
momentum distribution is found to be anisotropic, with the deformation
accentuated in the surface region of the finite Fermi system.

To incorporate both of these missing effects into the DME, Gebremariam
et al.\ have
constructed a model for the local momentum distribution based on previous
studies of the Wigner distribution function in nuclei \cite{Biruk}. 
The model parameters are adjusted so that the DME accurately
reproduces both integrated quantities, such as the expectation value of the
finite-range nucleon-nucleon interaction taken between Slater determinants
from self-consistent Skyrme-Hartree-Fock calculations, as well as the density
matrices themselves. The improvements are substantial, typically reducing
relative errors in integrated quantities by as much as an order of
magnitude across many
different isotope chains.  The improvement is especially striking for the
vector density matrices. 
We will test this improved DME in future investigations.

The tests of the DME will include benchmarks
against \textit{ab initio} methods in the overlap region
of light-to-medium nuclei.
Additional information is obtained from putting 
the nuclei in external fields,
which can be added directly to the DFT/DME functional. 
Work is in progress on comparisons to both 
coupled cluster and full configuration interaction calculations.
A key feature is that we use the same Hamiltonian for the microscopic
calculation and the DME approximation to the DFT.
The freedom to adjust (or turn off) external fields as well as to vary
other parameters in the Hamiltonian permits detailed evaluations of
the approximate functionals. 
In parallel there will be
refined nuclear matter calculations;
power counting arguments from re-examining the Brueckner-Bethe-Goldstone
approach in light of low-momentum potentials
will provide a framework for organizing higher-order contributions.
These investigations should provide insight into how
the energy density functional can be fine tuned for greater
accuracy in a manner
consistent with power counting and EFT principles.


\begin{ack}
We thank J.~Drut,
T.~Duguet, J.~Engel, B.~Gebremariam, N.~Kaiser, R.~Perry, V.~Rotivale,
and A. Schwenk for useful discussions. 
This work was supported in part by the National Science 
Foundation under Grant Nos.~PHY--0354916, PHY--0653312, PHY--0758125,
and PHY--0456903,  
and the UNEDF SciDAC Collaboration under DOE Grant 
DE-FC02-07ER41457.
\end{ack}


\end{document}